\documentclass[reprint, amsmath, amssymb, one column, aps]{revtex4-2}

\usepackage{graphicx}
\usepackage{subfigure}
\usepackage{subfloat}
\usepackage{amssymb}
\usepackage{amsmath}
\usepackage{multirow}
\usepackage{dcolumn}
\usepackage{bm}
\usepackage{wasysym}
\usepackage{cancel}

\usepackage{txfonts}
\usepackage{lineno}

\begin{document}

\title{Connecting Dynamo Theory with DNS Data: A Computational Analysis of $\alpha$ and $\beta$ Effects}

\author{Kiwan Park}
 \affiliation{Institute of Plasma Turbulence and Magnetic Fields, 50, Yongdam-ro, Sangdang-gu, Cheongju-si, Chungcheongbuk-do, Republic of Korea, 28717\\ pkiwan@iptm.re.kr, pkiwan@gmail.com\\}

\date{\today}


\begin{abstract}

We investigate the influence of current helicity on the turbulent magnetic diffusivity $\beta$ using three complementary derivations of the $\alpha$ and $\beta$ coefficients, based on the large-scale magnetic field $\overline{\mathbf{B}}$, the turbulent velocity $\mathbf{u}$, and the turbulent magnetic field $\mathbf{b}$. Applying these coefficients to raw DNS data, we reconstruct $\overline{\mathbf{B}}$ and compare the results with the original simulations. In the kinematic regime all models agree well with the DNS data. In the nonlinear regime, however, $\beta_{\mathrm{vv-vw}}$ alone produces unbounded growth of $\overline{\mathbf{B}}$. Including the contribution from turbulent magnetic fields ($\beta_{\mathrm{bb+jb}}$) suppresses this unphysical growth and restores agreement with the DNS results. We find that kinetic helicity drives $\beta$ more negative, while current helicity shifts it back toward zero. Weighted combinations of the coefficients further show that the $\beta$ effect dominates the evolution of $\overline{\mathbf{B}}$ throughout, whereas the $\alpha$ effect becomes important mainly for sustaining the field in the nonlinear regime. The corresponding IDL analysis scripts are provided to facilitate practical implementation of the theoretical models.
\end{abstract}


\maketitle

\section{Introduction}
Macroscopic astrophysical plasma systems are described by the scalar density $\rho$, polar velocity $\mathbf{U}$, and axial magnetic field $\mathbf{B}$, which form the foundamental variables of magnetohydrodynamics (MHD, \cite{2005PhR...417....1B}, \footnote{$\nu$ and $\eta$ indicate kinematic viscosity and molecular magnetic diffusivity.}):
\begin{eqnarray}
\frac{\partial \rho}{\partial t}&=&- {\bf \nabla}\cdot ({\bf \rho U})\label{continuity equation_original}\\
\frac{\partial {\bf U}}{\partial t}&=&-{\bf U} \cdot {\bf \nabla}\mathbf{U}-{\bf \nabla} \mathrm{ln}\, \rho + \frac{1}{\rho}{\bf J}\times {\bf B}+\nu\big({\bf \nabla}^2 {\bf U}+\frac{1}{3}{\bf \nabla} {\bf \nabla} \cdot {\bf U}\big)\label{momentum equation_original}\\
\frac{\partial \mathbf{B}}{\partial t}&=&\nabla \times (\mathbf{U}\times \mathbf{B}) +\eta \nabla^2\mathbf{B}.
\label{magnetic induction equation_original}
\end{eqnarray}
These nonlinearly coupled equations can be solved either numerically or analytically under restricted conditions \cite{2005PhR...417....1B, 2008ApJS..178..137S, 2021JOSS....6.2807P}). In addition, in rotating plasma structures such as solar and stellar magnetic fields, geomagnetic and planetary fields, galactic magnetic fields, protoplanetary disks, or AGN jets, the (large scale) magnetic induction equation Eq.~(\ref{magnetic induction equation_original}) can be simplified using $\alpha$ and $\beta$ tensors \cite{2002PhRvL..89z5007B}. Buoyancy and Coriolis forces in the rotating plasma systems give rise to kinetic helicity, defined as $\langle \mathbf{U} \cdot \nabla \times \mathbf{U} \rangle(\equiv H_V)$. Kinetic helicity in turn generates magnetic helicity, $\langle \mathbf{A} \cdot \mathbf{B} \rangle(\equiv H_M)$, with $\mathbf{B} = \nabla \times \mathbf{A}$. These pseudoscalars—arising from the interaction of an axial vector and a polar vector—together with the kinetic and magnetic energies, contribute to the $\alpha$ and $\beta$ effects, which govern the evolution of magnetic fields in the helical plasma systems:
\begin{eqnarray}
\frac{\partial \overline{\mathbf{B}}}{\partial t}=\nabla \times \underbrace{\langle \mathbf{u}\times \mathbf{b} \rangle}_{EMF}+\eta \nabla^2\overline{\mathbf{B}} \sim \alpha  \underbrace{\nabla \times \overline{\mathbf{B}}}_{\mathbf{J}}+(\beta+\eta) \nabla^2\overline{\mathbf{B}}.
\label{magnetic induction equation_alpha_beta}
\end{eqnarray}


This equation illustrates the mechanism by which magnetic fields are amplified or dissipate in plasmas. In plasmas made of many charged particles, the magnetic field evolves through both current density $\bf J$ (Ampere's law) and diffusion \citep{1999GMS...111..301B, 2003dysu.book..217P, 2005PhR...417....1B, 2007ApJ...658..129P, 2011AnRFM..43..583J, Akira2011}. Even when other physical drivers such as differential rotation ($d\Omega/dr$), shear, or magnetorotational instability (MRI) exist, the evolution of magnetic fields basically depends on electromagnetic effects and diffusion, especially in rotating plasma systems. Thus, if the $\alpha$ and $\beta$ values are calculated accurately, the evolution of magnetic fields nonlinearly coupled with plasma and their back-reaction effects can be explained and predicted more intuitively. Moreover, further extension and deepening related models can be expected.\\


The first conceptual $\alpha$ effect was proposed by Parker \cite{1955ApJ...122..293P}. He suggested that buoyancy lift the toroidal magnetic flux, $\mathbf{B}_{\text{tor}}$, and that the Coriolis force twist the rising magnetic tube by an angle of approximately $\pi/2$ through the $\alpha$ effect, resulting in a rotated magnetic loop and the generation of poloidal magnetic flux, $\mathbf{B}_{\text{pol}}$. However, Parker's $\alpha$ effect was developed through a combination of semi-empirical and theoretical approaches to explain observed phenomena. Consequently, the terminology and notation used are not fully consistent with those of modern dynamo theory. More rigorous theoretical frameworks—such as mean-field theory (MFT), the eddy-damped quasi-normal Markovian (EDQNM) approximation, and the direct interaction approximation (DIA)—have been employed to calculate these coefficients \citep{1966ZNatA..21.1285S, 1976JFM....77..321P, 1978mfge.book.....M, 1980opp..bookR....K, Akira2011}. Besides, approaches such as path-integral method \citep{2025ApJ...985...18R} and renormalization group theory (RNG, \citep{2023PhRvE.107e5205M}) have been applied to turbulence calculations; however, their applicability to HD or MHD requires further detailed verification. Moreover, applying these models to observations or experiments remains a highly nontrivial task.\\

The formal expressions of $\alpha$ and $\beta$ vary depending on the closure model used. $\alpha$ generally depends on the turbulent residual helicity, $\alpha\sim \tau \big(\langle \mathbf{j} \cdot \mathbf{b} \rangle - \langle \mathbf{u} \cdot \bm{\omega} \rangle\big)$, which is considered responsible for the amplification of the large scale magnetic field $\overline{\mathbf{B}}$ ($\langle \mathbf{j} \cdot \mathbf{b} \rangle \neq \langle \mathbf{u} \cdot \bm{\omega} \rangle$ or its saturation ($\langle \mathbf{j} \cdot \mathbf{b} \rangle \sim \langle \mathbf{u} \cdot \bm{\omega} \rangle$). In contrast, $\beta\sim\tau\langle u^2\rangle$ (or $\tau\langle b^2\rangle$) is considered to diffuse $\overline{\mathbf{B}}$. Strictly speaking, these inferences are based on the partial characteristics of the leading terms composing $\alpha$ and $\beta$. For instance, the $\alpha$-effect has been widely regarded as the primary driver of magnetic field amplification. And, theoretical models have also been developed along these lines. However, in a plasma where opposing charges largely neutralize one another, asserting that the electromagnetic $\alpha$-effect plays the dominant role is rather a oversimplification: $\partial {{\bf B}} /\partial t \sim \alpha {\bf J}$ (${J}={J_1}\pm{J_2}\pm{J_3}+...$).\\

On the other hand, $\beta$ has been regarded as a positive scalar, diffusing the magnetic field according to $\beta \nabla^2 {\mathbf{B}} \rightarrow - \beta k^2 {\mathbf{B}}$. The form of magnetic diffusion dependent on the Laplacian $\nabla^2$ shares the same mathematical structure as fluid diffusion. However, plasma diffusion is not so straightforward, as the Lorentz-force interactions of electromagnetic eddies can potentially generate current density. The induction of current density exerts a negative effect on $\beta$; however, these dynamics have not yet been sufficiently studied.\\

Although not the mainstream view, the possibility of a negative $\beta$ has been persistently proposed. Moffatt\citep{1974JFM....65....1M} derived the representations of $\alpha$ and $\beta$ in a Lagrangian formulation, and Kraichnan \citep{1976JFM....75..657K} rederived the magnetic induction equation in a strongly helical system: $\partial_t \overline{\mathbf{B}} \sim \tau_2 \nabla \times \langle \alpha \nabla \times \alpha \rangle \, \overline{\mathbf{B}} \sim -\tau_2 A \nabla^2 \overline{\mathbf{B}},\, \beta_{turb}\rightarrow-\tau_2A<0$, where $\langle \alpha \alpha' \rangle = A(x-x')\,D_2(t-t')$ and $\tau_2 = \int^{\infty} D_2(t)\, dt$. This result based on the correlations between $\alpha(x,\,t)$ and $\alpha(x',\,t')$, implicitly assumes the long-term stability of the helical field and the memory effect ($\sim \tau_2$) in large scale eddies. Moreover, in recent studies, Rogachevskii et al.\ \citep{2025ApJ...984...88B, 2025ApJ...985...18R} proposed a turbulent magnetic diffusion coefficient, $\eta_t$, derived using path integrals, and demonstrated that magnetic diffusivity qualitatively decreases in the presence of kinetic helicity. \\

From a numerical perspective, the Test-Field Method (TFM) has been developed to extract the $\alpha$ and $\beta$ coefficients from simulation data by adding an additional test field to the code \citep{2005AN....326..245S, 2025ApJ...984...88B}. The results show that the $\beta$ effect is suppressed by kinetic helicity. More recently, Bendre et al.\ \citep{2024MNRAS.530.3964B} introduced a numerical method based on the Högbom CLEAN algorithm and the iterative removal of sources (IROS) technique, which employs the time series of the mean magnetic field and current as inputs. Their results also show the partially negative $\beta$ diffusivity. Furthermore, the negative magnetic diffusivity has been observed in liquid sodium experiments \citep{2014PhRvL.113r4501C, 2014NJPh...16g3034G, 2015ApJ...811..135A}. Small-scale turbulent fluctuations were shown to contribute to this effect in the interior region.\\

For practical purposes, considerable efforts have been made to incorporate the $\alpha$ and $\beta$ coefficients into solar dynamo models \cite{2003dysu.book..217P, 2014ARA&A..52..251C}, and similar approaches have also been explored in the context of the geodynamo \cite{2005AN....326..245S}. These efforts primarily aim to reproduce the observed spatial structure and periodicity of the evolving solar magnetic field using $\alpha$ and $\beta$ coefficients extracted from direct numerical simulations (DNS). \citet{2013ApJ...768...16S} obtained the $\alpha_{ij}$ tensor by combining the numerical technique introduced by \citet{2011ApJ...735...46R} with the theoretical framework of \citet{1978mfge.book.....M}, employing the relation $\boldsymbol{\xi}_i = \langle \mathbf{u} \times \mathbf{b} \rangle_i = \alpha_{ij} \overline{B}_j + \beta_{ijk} \partial_k \overline{B}_j$. Since Racine et al. focused on the $\alpha$ tensor, Simard et al. used an inferred form of simplified $\beta$ to represent turbulent diffusivity, and adopted representative parameter values such as $\alpha_0$, $\eta_0$, and $\Omega_0$. Their approach can be compared with that of \citet{2008A&A...483..949J}, who determined $\alpha$ through a trial-and-error procedure to ensure numerical stability while reproducing the large-scale features of the solar magnetic field.\\ 


To examine the roles of $\alpha$ and $\beta$ more directly, we reconstruct the magnetic fields by applying weighting factors to the verified $\alpha$ and $\beta$ profiles and compare them with the DNS results. To show how this theoretical model is implemented and to help readers reproduce the results, the IDL analysis code is provided in the Appendix. \\

We adhered to straightforward and robust methods to prioritize an intuitive understanding of the underlying theory. The code has been utilized in this study and verified for proper execution. Should readers encounter any execution errors, the corresponding author may be contacted directly to receive the source files.\\

It should be emphasized that this manuscript focuses on bridging the theoretical framework with its computational implementation. Since Parker introduced the concept of the dynamo, numerous theoretical models aiming to determine alpha ($\alpha$) and beta ($\beta$) have been proposed. However, these efforts have continually introduced new models without revealing the exact solutions or profiles. In this context, constructing theoretical models for $\alpha$ and $\beta$ through two distinct approaches and applying them to actual analysis codes carries significant importance.Most importantly, $\alpha$ and $\beta$ are no longer abstract concepts but physical quantities that can be physically identified, providing a framework to expand and deepen the theory of rotating plasma astrophysics. Furthermore, incorporating these theoretical models into a kinetic Boltzmann model allows for the direct observation of how turbulent motion affects particles through the $\alpha$ and $\beta$ profiles.In plasma physics, most magnetic field studies rely on fixed external background fields, while dynamic magnetic field applications remain rare. Therefore, attempting to analyze $\alpha$ and $\beta$ under these dynamic conditions is highly timely.\\

This paper is organized as follows. Section 2 introduces the calculation of $\alpha_{EM\text{-}HM}$ and $\beta_{EM\text{-}HM}$ from large-scale magnetic field data, together with the corresponding simulation results. Section 3 revisits the method of deriving $\beta_{vv-vw+bb+jb}$ from kinetic and magnetic energy and helicity, and compares the reconstructed large-scale magnetic field with the DNS results. To explain the suppressed convergence of the magnetic field in the nonlinear regime, turbulent magnetic energy and helicity are newly introduced, and the resulting magnetic field is compared with DNS. Additionally, the corresponding IDL source code will be described. In addition, the direct influence of $\alpha$ and $\beta$ on dynamo action is systematically investigated by introducing weighting parameters to these coefficients. The last section provides a summary. Finally, the full IDL source code is provided in the Appendix, and the Pencil Code dataset has been uploaded to facilitate learning.
And, they are available in the Github repository, \url{https://github.com/pkiwan93/dynamo-dns-analysis}. The Pencil Code simulations were done using only the standard, built-in features, which have been verified by many users and developers. Additionally, the kinetic helicity acting as the driving source takes opposite signs in the Northern and Southern Hemispheres of a rotating body. While $+1$ for the Southern Hemisphere is conventional, we used $-1$ for the Northern Hemisphere.


\begin{figure*}[t]
\centering
\subfigure[$\int E_V(k) dk =  \langle U^2\rangle/2$, $\int E_M(k) dk =  \langle B^2\rangle/2$]{
    \includegraphics[width=0.46\textwidth]{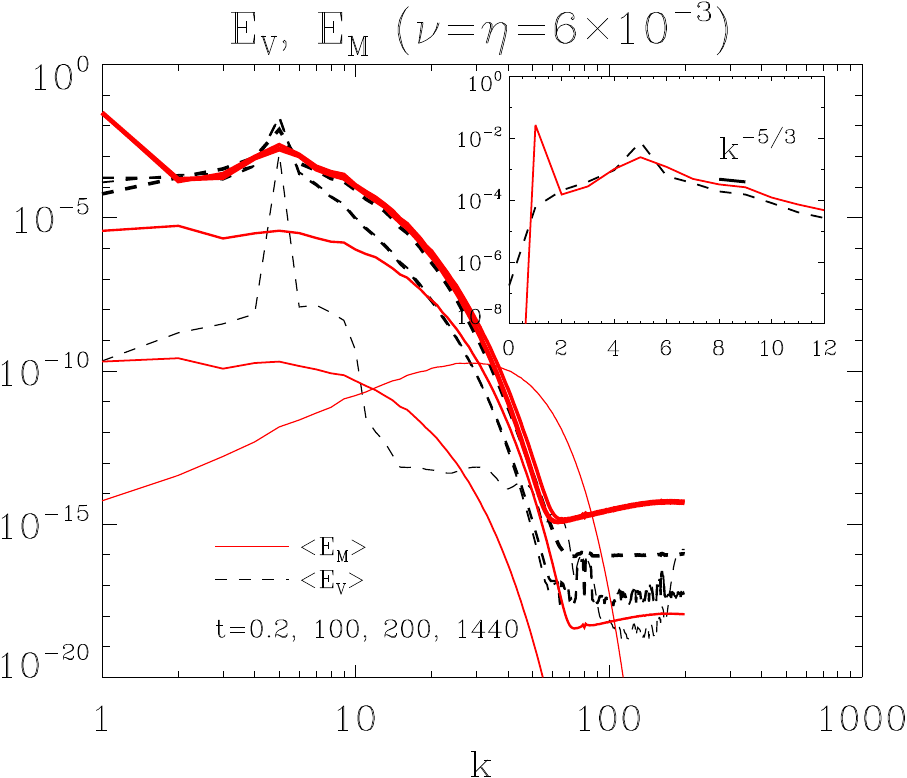}
    \label{f1}
}\hspace{-3 mm}
\subfigure[$H_C(k,t) (\equiv \langle {\bf j}\cdot {\bf b} \rangle$) =$k^2 H_M(k,t)$, $H_V<0$]{
    \includegraphics[width=0.46\textwidth]{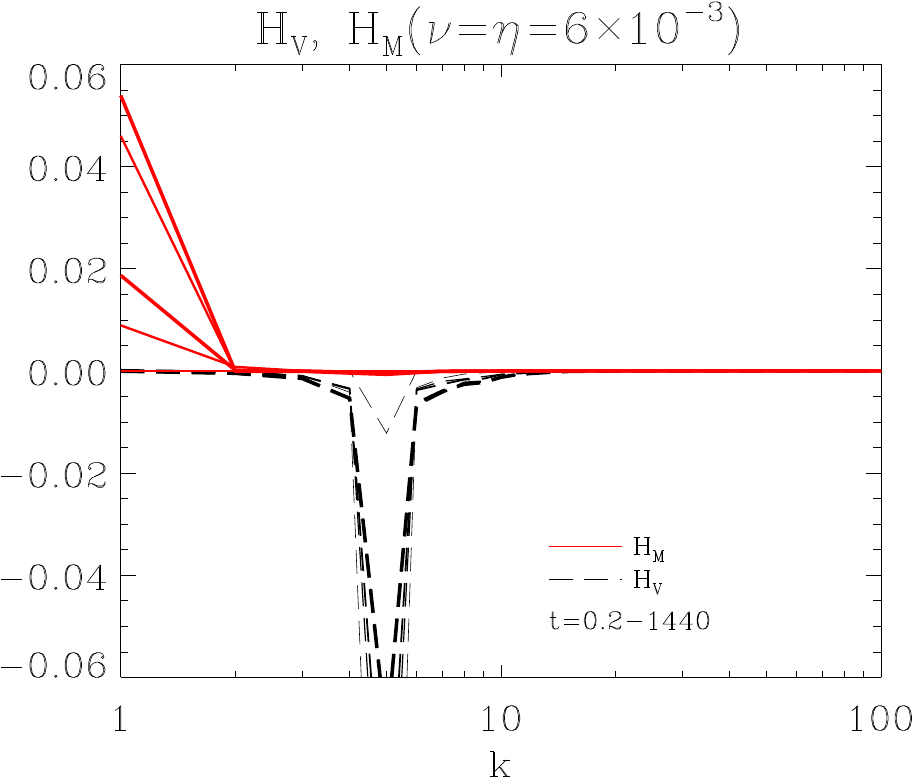}
    \label{f1a}
}\hspace{-3 mm}
\subfigure[$\alpha(t)$]{
    \includegraphics[width=0.46\textwidth]{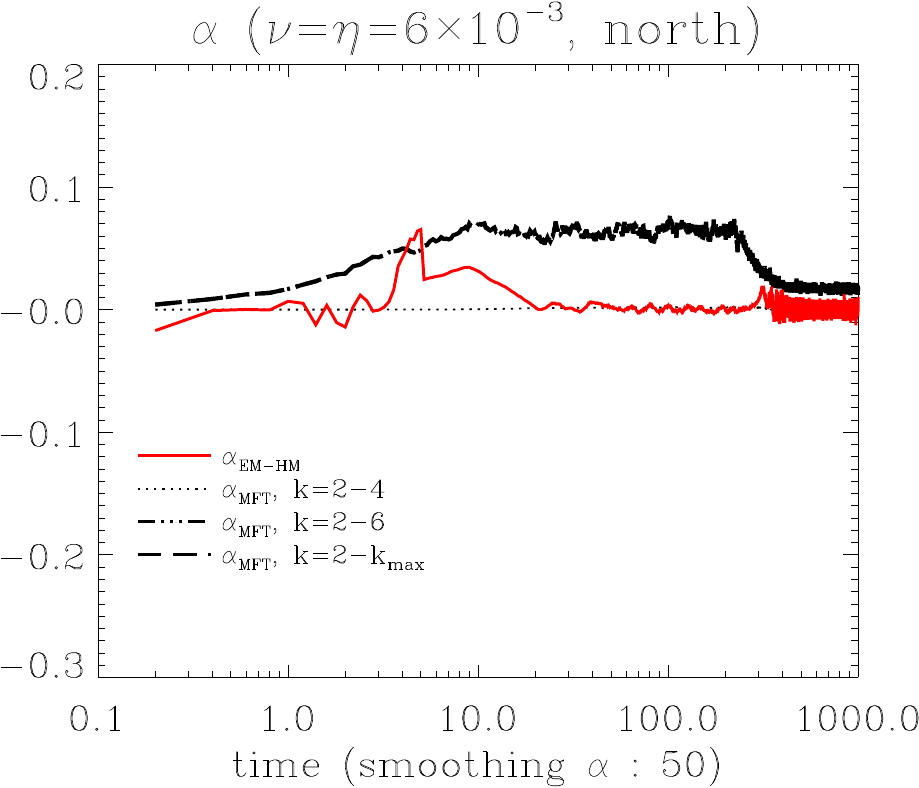}
    \label{f2}
}\hspace{-3 mm}
\subfigure[$\beta(t)$]{
    \includegraphics[width=0.46\textwidth]{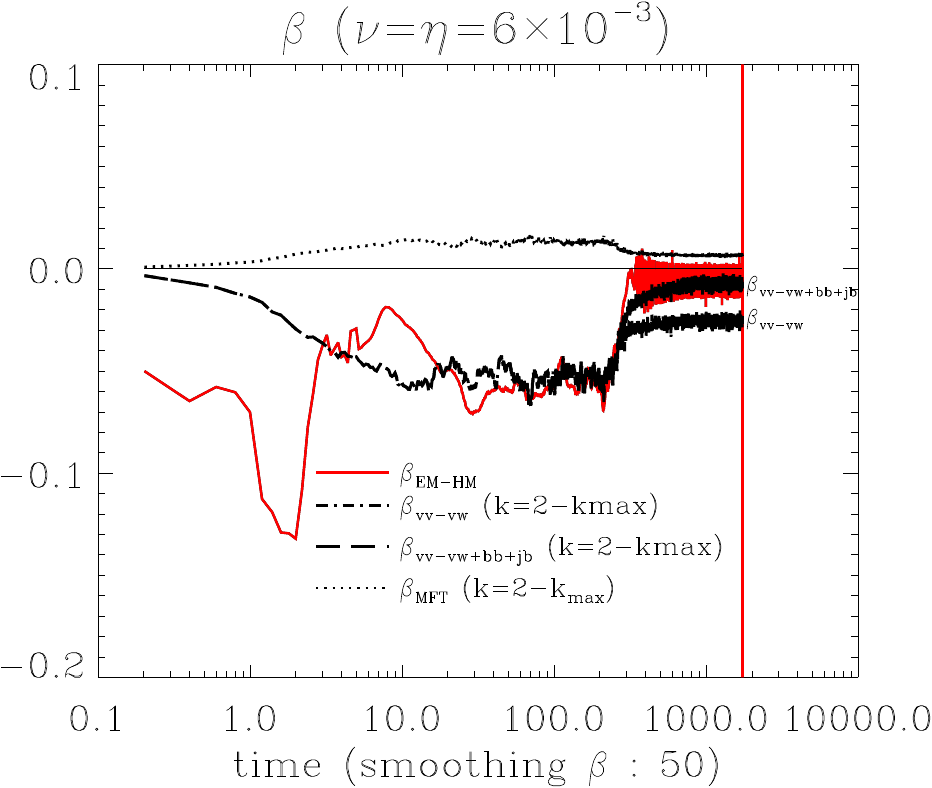}
    \label{f3}
}
\caption{Negative helical kinetic forcing at the eddy scale $k=5$, representing the Northern Hemisphere of the plasma. Raw MHD data are applied to calculate $\alpha$ and $\beta$ within the forcing scale regime ($k=2$–$9$).  \textit{Reassembled and adapted from Ref.~\cite{particles8040098}. This comparison provides the baseline for the newly integrated IDL analysis code and extended theoretical formulations presented in this study.}}
\end{figure*}

\begin{figure*}[t]
\centering
\subfigure[$\nabla\times \langle \mathbf{u}\times \mathbf{b} \rangle$]{
    \includegraphics[width=0.46\textwidth]{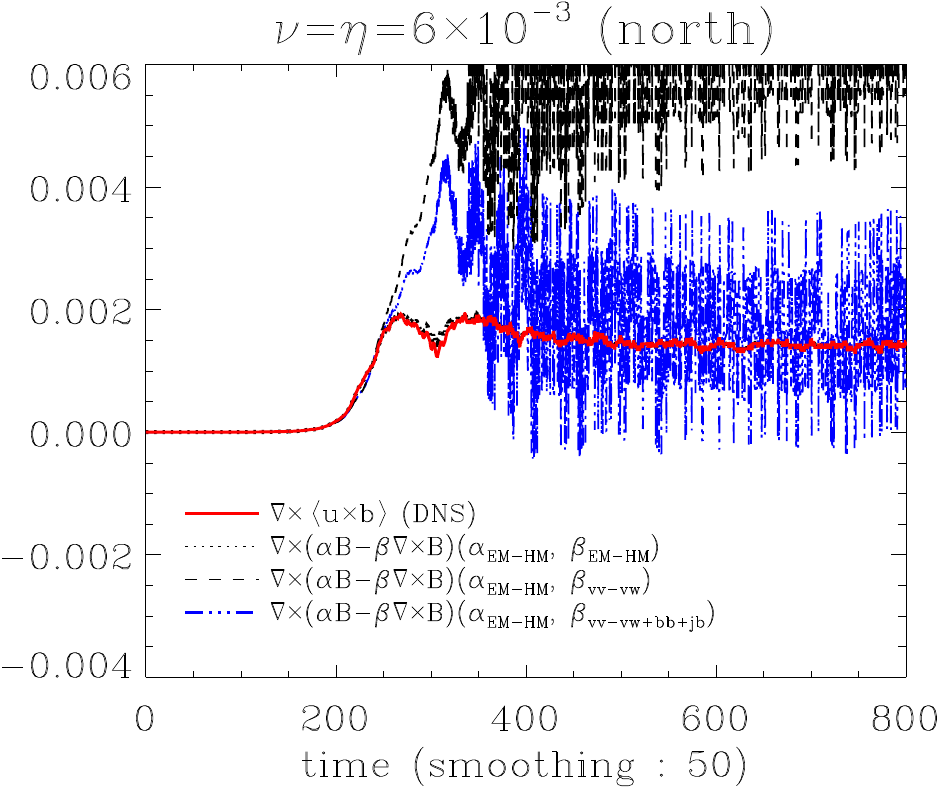}
    \label{f3a}
}\hspace{-3 mm}
\subfigure[$\overline{B}(t)$ (k=2-9)]{
    \includegraphics[width=0.46\textwidth]{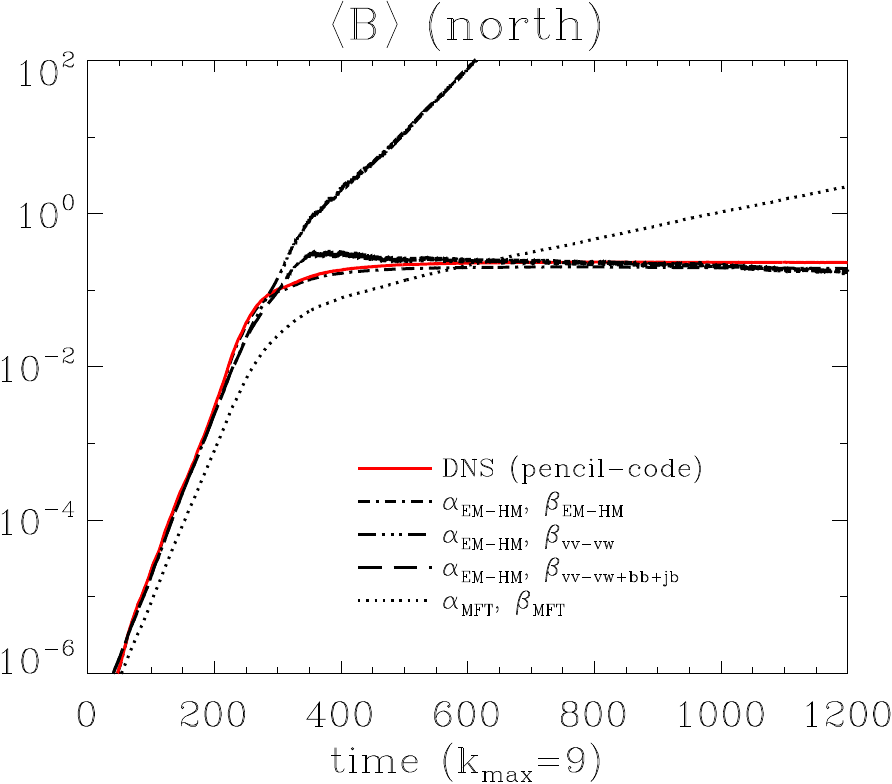}
    \label{f4}
}
\caption{(a) The forcing scale regime $k=2$–$9$ is analyzed. For $k=2$–$8$ and $k=2$–$200$, see Figures A1(b), A1(e), and A1(f) in the Appendix.   (b) \textit{[Adapted from Ref. ~\cite{particles8040098}, with the inclusion of the large-scale magnetic helicity ratio to enhance precision.]}}
\end{figure*}

\begin{figure*}[t]
\centering
\subfigure[$\overline{B}(t)$ with varying $\alpha$]{
    \includegraphics[width=0.46\textwidth]{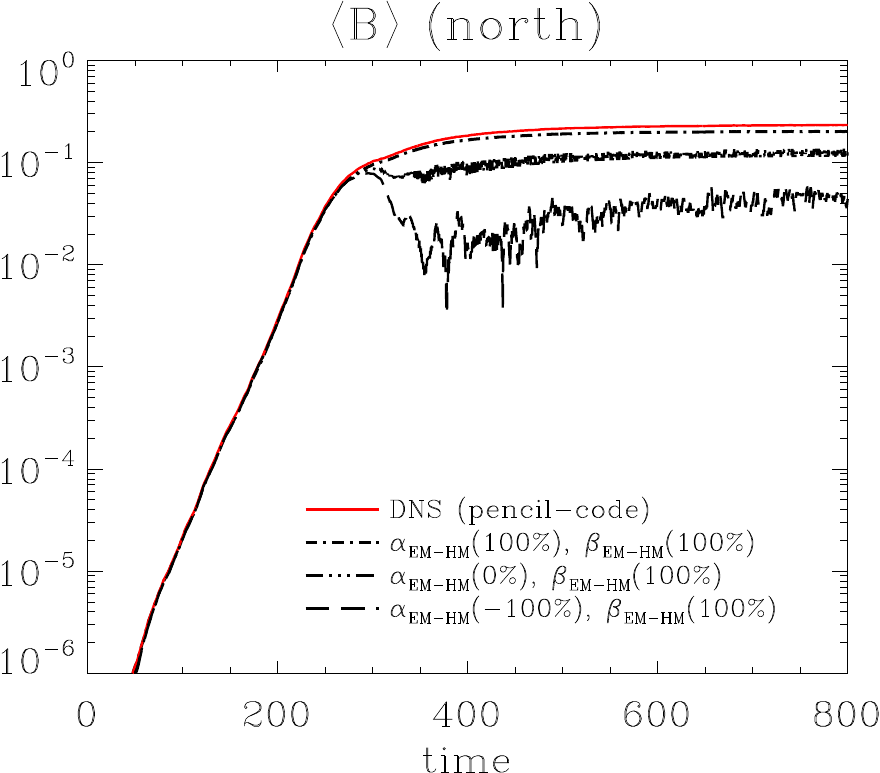}
    \label{f5}
}\hspace{-3 mm}
\subfigure[$\overline{B}(t)$ with varying $\beta$]{
    \includegraphics[width=0.46\textwidth]{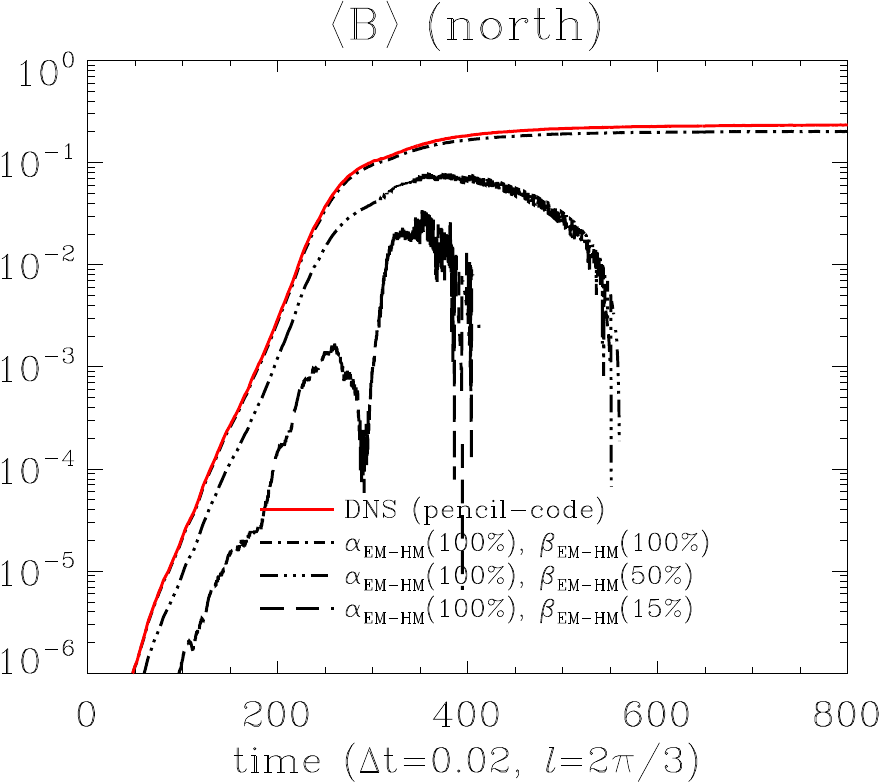}
    \label{f6}
}\hspace{-3 mm}
\subfigure[$\overline{B}(t)$ with varying $\alpha$]{
    \includegraphics[width=0.46\textwidth]{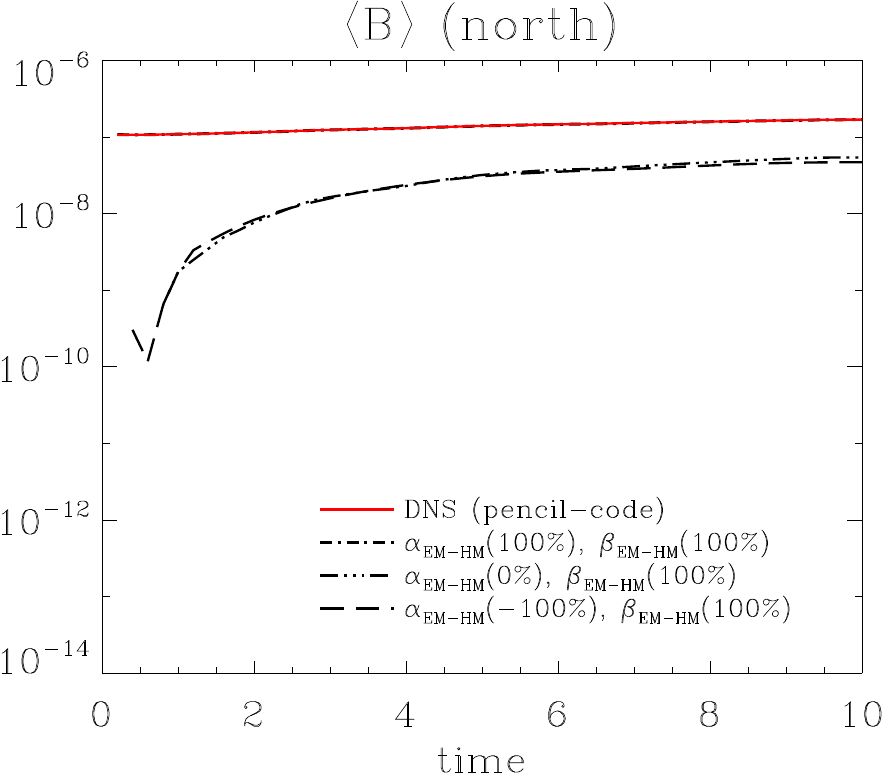}
    \label{f7}
}\hspace{-3 mm}
\subfigure[$\overline{B}(t)$ with varying $\beta$]{
    \includegraphics[width=0.46\textwidth]{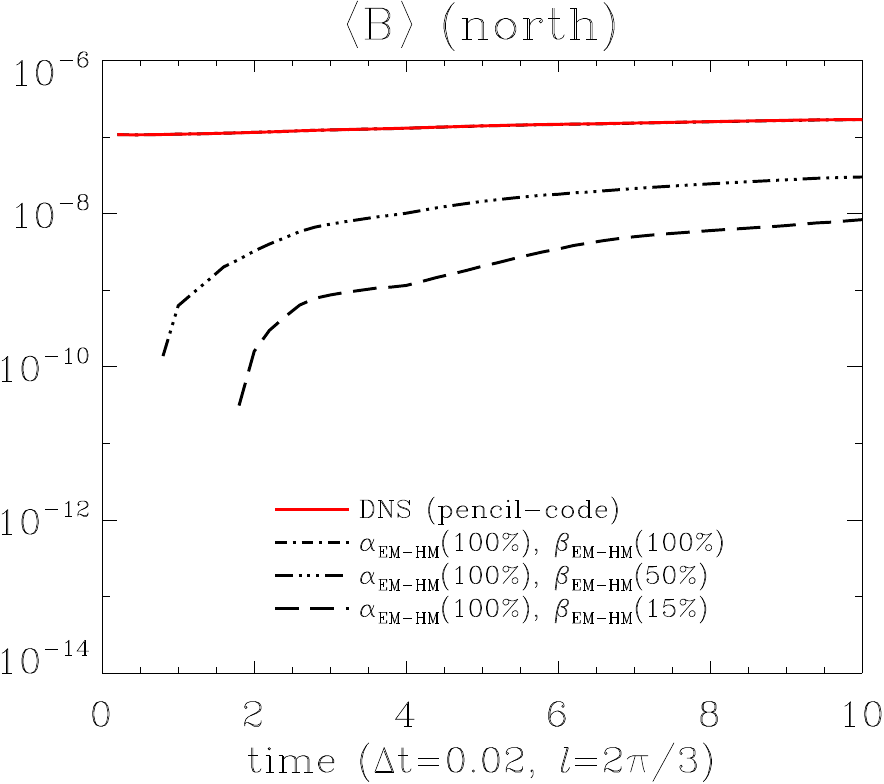}
    \label{f8}
}
\caption{(a) $\alpha$-effect: $-100\%$, $0$, $100\%$; $\beta$-effect: $100\%$. (b) $\alpha$-effect: $100\%$; $\beta$-effect: $15\%$, $50\%$, $100\%$. (c) Same as (a) for the early-time regime. (d) Same as (b) for the early-time regime.}
\end{figure*}

\section{Analytical Model and Numerical Validation}
\subsection{Derivations of $\alpha_{EM-HM}$ and $\beta_{EM-HM}$ and results}

In our prior research, we derived $\alpha$ and $\beta$ from large-scale magnetic energy $\overline{E}_M( =\langle \overline{B}^2\rangle/2)$, and magnetic helicity $\overline{H}_M( =\langle \overline{A}\cdot \overline{B})\rangle$.
\begin{eqnarray}
\frac{\partial \overline{H}_M}{\partial t} &=&  4\alpha \overline{E}_M - 2(\beta + \eta)\overline{H}_M, \label{Hm1} \\
\frac{\partial \overline{E}_M}{\partial t} &=&  \alpha \overline{H}_M - 2(\beta + \eta)\overline{E}_M. \label{Em1}
\end{eqnarray}
By changing bases, we obtain
\begin{eqnarray}
\alpha&=&\frac{1}{4}\frac{d}{dt}log_e \bigg|\frac{ 2\overline{E}_M(t)+\overline{H}_M(t)}{2\overline{E}_M(t)-\overline{H}_M(t)}\bigg|\sim \frac{\Delta log_e[\cdot]}{4\Delta t},\label{alphaSolution3}\\
\beta&=&-\frac{1}{4}\frac{d}{dt}log_e\big| \big(2\overline{E}_M(t)-\overline{H}_M(t) \big)\big( 2\overline{E}_M(t)+\overline{H}_M(t)\big)\big|-\eta. \label{betaSolution3}
\end{eqnarray}
$\alpha(\rightarrow \alpha_{EM-HM})$ and $\beta(\rightarrow \beta_{EM-HM})$ are functions of the pseudoscalar $\overline{H}_M(t)$, the true scalar $\overline{E}_M(t)$, and the time interval $\Delta t$. Under reflection symmetry, $\overline{E}_M(t)$ remains invariant, whereas $\overline{H}_M(t)$ changes sign. Consequently, the numerator and denominator are interchanged, leading to a sign reversal of $\alpha$, consistent with its pseudoscalar nature. In contrast, $\beta$ in Eq.~(\ref{betaSolution3}) is unaffected by reflection and thus retains its scalar character. \footnote{Arithmetic operations are performed between the scalar $\overline{E}_M$ and the pseudoscalar $\overline{H}_M$. Although operations involving scalars, pseudoscalars, axial vectors, and polar vectors appear to be prohibited, they are allowed provided that the physical symmetries are preserved. All forms of helicity inherently constitute pseudoscalars through axial--polar vector interactions, much like the Lorentz force, ${\bf u}\times{\bf b}$, and the Poynting vector arise from cross products involving different types of vectors. Moreover, relativistic four-vectors, as well as MHD and Boltzmann kinetic models, combine distinct tensor types within a single set of governing equations. Basic operations such as addition and subtraction are possible as long as the relevant physical properties are preserved. Here, applying reflection symmetry preserves each coefficient's own parity: $\alpha$ remains pseudoscalar, $\beta$ remains true scalar.}\\

The raw data of $\overline{H}_M$ and $\overline{E}_M$ can be obtained from standard MHD codes or observational data sets if available. In this work, we employed DNS data from Pencil Code \citep{2021JOSS....6.2807P} in a $(2\pi)^3$ cube system, with both $\nu$ and $\eta$ set to 0.006, applying fully helical kinetic forcing of amplitude $f_0 = 0.07$ at the forcing scale $k_f = 2\pi/l = 5$. In a three-dimensional box system with each side of length $2\pi$, $k=1$ corresponds to the large scale. A more general representation $k_1$ can be adopted. Then, the coupled equations can be solved by changing bases. However, this introduces unnecessary complexity into the dimensionless MHD model with the box size of $(2\pi)^3$ ($k_1=2\pi / 2\pi$). Therefore, we retain the simpler choice $k=1$. For further details, the reader may refer to the Pencil Code manual. We carried out parallel computations on 80 CPUs with a grid resolution of $400^3$ and analyzed the resulting data using IDL. We can now obtain accurate profiles of $\alpha$ and $\beta$ as they evolve over time and under varying physical conditions by replacing the theoretically indeterminable fundamental physical variables with DNS raw data.\\

\noindent Then, $\alpha$ and $\beta$ in Eqs.~(\ref{alphaSolution3}), (\ref{betaSolution3}) are calculated as follows:
\begin{verbatim}
for j=0L,  (i_last-1) do begin

    c[j]=2.0*spec_mag(1, j) + spechel_mag(1, j) 
    d[j]=2.0*spec_mag(1, j) - spechel_mag(1, j)  
 
endfor

for j=0L,  i_last-1 do begin

    alpha[j] =0.25*( (ALOG(c[j+1])-ALOG(c[j]) )-(ALOG(d[j+1])-ALOG(d[j])) ) $
    /( tt_power_mag_time[j+1]-tt_power_mag_time[j] )

    beta[j] =-0.25*( (ALOG(c[j+1])-ALOG(c[j]))+(ALOG(d[j+1])-ALOG(d[j])) ) $
    /( tt_power_mag_time[j+1]-tt_power_mag_time[j] )-eta

endfor
\end{verbatim}

\noindent \verb|spec_mag(1, j)| and \noindent\verb|spechel_mag(1, j)| indicate $\overline{E}_M(t)$ and $\overline{H}_M(t)$. The number '1' means the large scale $k=1$ in Fouier space. ALOG denotes the natural logarithm function in IDL. Also, the value $0.99$ is used to avoid division by zero in the denominator when the magnetic field reaches a fully helical state: $2\overline{E}_M\sim\pm\overline{H}_M$.\\

Figure\ref{f1} shows the spectra of kinetic energy $E_V$ (black dashed line) and magnetic energy $E_M$ (red solid line) from DNS at $t = 0.2$, 100, 200, and 1440. The vertical axis denotes the energy density, while the horizontal axis denotes the wavenumber $k$. The figure illustrates a typical inverse energy cascade, in which the magnetic energy at the large scale ($k=1$) exceeds the kinetic energy at the forcing scale. The inset highlights the energy density in the range $k=1$–12 at $t=1440$, when both $E_M$ and $E_V$ have saturated, revealing the emergence of a short Kolmogorov inertial range between $k=8$ and 9. This range corresponds to the typical forcing-scale regime.\\ 

Figure \ref{f1a} shows $H_M(k, t)$ and $H_V(k, t)$ in Fourier space. Since it represents the Northern Hemisphere, a left-handed $H_V$ (black dashed line) is given at $k=5$, and the resulting $H_M$ (red solid line) exhibits right-handedness at large scales. The data in the Figures \ref{f1}, \ref{f1a} determine $\alpha$ and $\beta$.\\

Figure \ref{f2} compares $\alpha_{\mathrm{EM\text{-}HM}}$, obtained using Eq.~(\ref{alphaSolution3}), with $\alpha_{\mathrm{MFT}}\sim 1/3 \tau \int^k (\langle \mathbf{j}\cdot \mathbf{b}\rangle-\langle \mathbf{u}\cdot\nabla\times \mathbf{u}\rangle)\, dk$, derived from mean-field theory. For visual clarity, we averaged over 50 neighboring data points. This originates from the Pencil Code scheme, where values are randomly chosen keeping $\langle k_f\rangle$. While we used a smoothing technique in IDL, other filters like FFT can also be applied. The value of $\alpha_{\mathrm{MFT}}$ remains positive and converges to zero in the saturated regime of the magnetic field, which underlies the conventional inference of the $\alpha$ profile. In contrast, $\alpha_{\mathrm{EM\text{-}HM}}$ exhibits initial oscillations and approaches zero much more rapidly, implying that the role of $\alpha$ in the dynamo is limited. In plasma, particles with opposite charges counteract one another, thereby restricting the electromagnetic contribution to magnetic field amplification. In southern hemisphere, where positive kinetic helicity drives the system, the sign of $\alpha$ reverses, but the sign of $\beta$ does not change. (see Figure. A\ref{fig:app_a} in Appendix)\\

\subsection{Derivations of $\beta_{vv-vw}$ and $\beta_{vv-vw+bb+jb}$ and results}

Eqs.~(\ref{alphaSolution3}), (\ref{betaSolution3}) yield the most accurate $\alpha$ and $\beta$ values among the models presented here. However, because they rely on an indirect approach, alternative methods such as a functional recursive method with statistical second order moment closure theory is necessary to clarify the underlying physical mechanism.\\

The theoretical derivation of $\beta$ begins with taking the time derivative of EMF as follows.
\begin{eqnarray}
\int^{\tau} \frac{\partial}{\partial t}\big\langle{\bf u}\times {\bf b}\big\rangle \, dt=\int^{\tau} \bigg(\bigg\langle\frac{\partial{\bf u}}{\partial t}\times {\bf b}\bigg\rangle+\bigg\langle{\bf u}\times \frac{\partial{\bf b}}{\partial t}\bigg\rangle\bigg) \, dt.
\label{beta_derivation1}
\end{eqnarray}
This expression follows from the derivative rule of a product, indicating that the time derivative of the turbulent velocity and that of the turbulent magnetic field act as independent sources. Both the linear kinematic regime (weak magnetic field) and the nonlinear regime (field amplification) are included to obtain a complete formulation. This procedure keeps the EMF intact while replacing each time-derivative term with the corresponding MHD equation, thereby yielding the $\alpha$ and $\beta$ coefficients.\\

In our earlier work \citep{2025PhRvD.111b3021P}, we found $\beta$ as follows:
\begin{eqnarray}
\bigg\langle \mathbf{u}\times \int^{\tau}\frac{\partial \mathbf{b}}{\partial t}dt \bigg\rangle&\rightarrow&
\tau\,\bigg\langle -\epsilon_{ijk} u_j (r)u_m(r+l)\frac{\partial \overline{B}_k}{\partial \overline{r}_m}\bigg\rangle\nonumber\\
&\sim&\underbrace{\tau\bigg(\frac{1}{3}\langle u^2\rangle - \frac{l}{6}H_V \bigg)}_{\beta_{vv-vw}}\big(-\nabla \times \overline{\bf B}\big).\label{beta_derivation2}
\end{eqnarray}
The second order moment $\langle u_j u_m \rangle$ gives rise to a two-point cross-correlation moment, $\langle u_j(r) u_m(r+l)\rangle$, which is expressed as \citep{2008tufl.book.....L}:
\begin{eqnarray}
\langle u_j(r)u_m(r+l)\rangle=\frac{\langle u^2\rangle}{3}\delta_{jm}-\epsilon_{jms}\frac{l_s}{6}H_V.
\label{uu_moment}
\end{eqnarray}
The Fourier-transformed form of this relation is 
\begin{eqnarray}
\frac{\langle u_i(k)u_j(k')\rangle}{\delta(k+k')}=E_V(k)\bigg(\delta_{ij}-\frac{k_ik_j}{k^2}\bigg)+\frac{ik_l}{2k^2}\epsilon_{ijl}H_V(k).
\label{uu_moment_in_FT}
\end{eqnarray}
Note that energy density $E_V=\langle u^2\rangle/2$ and kinetic helicity $H_V=\langle \mathbf{u}\cdot \bm{\omega}\rangle$ are single correlated moments. If $i=j$ or reflection symmetry is present, this relation gives rise to the conventional $\beta$ effect $\tau/3 \langle u^2\rangle$.\\

The velocity field is constrained by the Lorentz force; therefore, an additional term should be included for the magnetic back reaction $\mathbf{J}\times \mathbf{B}$. The turbulent magnetic tensions arising from the Lorentz force are $\overline{\mathbf{B}}\cdot\nabla \mathbf{b}$ and $\mathbf{b}\cdot\nabla\overline{\mathbf{B}}$. The former leads to the usual $\alpha$ effect, $\langle \int^{\tau} \partial \mathbf{u}/\partial t\, dt \times \mathbf{b} \rangle \sim \tau\langle \overline{\mathbf{B}}\cdot\nabla \mathbf{b} \times \mathbf{b} \rangle \sim \tau \langle \mathbf{b}\cdot \mathbf{j}\rangle\, \overline{\mathbf{B}}\sim \alpha \overline{\mathbf{B}}$, while the latter contribution can be expressed as follows:
\begin{eqnarray}
&&\bigg\langle \int^{\tau}\frac{\partial \mathbf{u}}{\partial t}dt \times \mathbf{b} \bigg\rangle \rightarrow
\tau\big\langle \mathbf{b}\cdot\nabla \overline{\mathbf{B}} \times \mathbf{b} \big\rangle\sim \tau\bigg\langle \epsilon_{ijk} b_m\frac{\partial \overline{B}_j}{\partial \overline{r}_m}b_k\bigg\rangle \nonumber\\
&&\sim\frac{\tau}{3}\langle b_k(r)b_m(r+l)\rangle \epsilon_{ijk}\frac{\partial \overline{B}_j}{\partial \overline{r}_m}\delta_{km}\nonumber\\
&&-\tau\big\langle \epsilon_{kms}\frac{l_s}{6}\big(\mathbf{b}\cdot \nabla\times \mathbf{b}\big)\big\rangle\epsilon_{ijk}\frac{\partial \overline{B}_j}{\partial \overline{r}_m}\,\,\,(m\rightarrow i,\,s\rightarrow j)\nonumber\\
%
%
&\rightarrow&-\frac{\tau}{3}\langle b^2\rangle \big(\nabla \times \overline{\bf B}\big)_i-\tau \frac{H_c}{6}\big(\bm{l}\times \big[\nabla\times \overline{\bf B}\big]\big)_i\,\,\,\nonumber\\
%
%
&\rightarrow&\tau\,\bigg(\frac{1}{3}\langle b^2\rangle + \frac{l}{6}H_C \bigg)\,\big(-\nabla \times \overline{\bf B}\big)\equiv \beta_{bb+jb} \, \big(-\nabla \times \overline{\bf B}\big).\label{general_beta_derivation_for_b8}
\end{eqnarray}
We have used turbulent magnetic energy and current helicity $H_C=\langle \mathbf{j}\cdot \mathbf{b}\rangle$ composed of an axial vector $\bf b$ and a polar vector $\bf j$:
\begin{eqnarray}
\langle b_k(r)b_m(r+l)\rangle=\frac{\langle b^2\rangle}{3}\delta_{km}-\epsilon_{kms}\frac{l_s}{6}H_C.
\label{general_beta_derivation5}
\end{eqnarray}
If $k = m$, the second term on the right hand side vanishes, and the expression reduces to the positive $\beta$ effect in DIA and EDQNM. If $k \neq m$ and reflection symmetry is broken, the complete form of $\beta$ becomes
\begin{eqnarray}
\beta_{vv-vw+bb+jb} = \int^{\tau}\bigg(\frac{2}{3}E_V + \frac{2}{3}E_M - \frac{l}{6}H_V + \frac{l}{6}H_C\bigg)\, dt.
\label{complete_beta_derivation}
\end{eqnarray}
Note that $lH_V$ and $lH_C$ do not change sign under reflection (in the code, reflection$\rightarrow -(-l)H_{V,\,C}$, the data of $H_V$ or $H_C$ change sign). The profiles of $\beta_{vv-vw+bb+jb}$ and the corresponding $\overline{\mathbf{B}}$ are shown in Figures \ref{f3} and \ref{f4} (dashed lines). Turbulent energies and current helicity enhance magnetic diffusivity (or suppress $\overline{\mathbf{B}}$), whereas kinetic helicity reduces diffusivity (or amplifies the field). Notably, the $\beta$ term also depends on the residual helicity effect, similar to the conventional $\alpha$ effect..\\

Figure \ref{f3} shows $\beta_{\mathrm{EM\text{-}HM}}$ (red solid line, Eq.~(\ref{betaSolution3})), $\beta_{\mathrm{MFT}}$ (dotted line, $\sim \tau/3 \int^k \langle u^2 \rangle\, dk$), $\beta_{vv-vw}$ (dot-dashed line, Eq.~(\ref{beta_derivation2})), and $\beta_{vv-vw+bb+jb}$ (dashed line, Eq.~(\ref{complete_beta_derivation})). $\beta_{\mathrm{MFT}}$ remains positive and decreases in the saturated regime, indicating the diffusion of $\overline{\mathbf{B}}$. In contrast, the other $\beta$ profiles remain negative and converge to zero. The profile of $\beta_{vv-vw}$ clearly demonstrates that the kinetic helicity significantly reduces diffusivity $\beta$ compared to $\beta_{\mathrm{MFT}}$ and coincides with $\beta_{\mathrm{EM-HM}}$ in the kinematic regime ($t < 250$). However, a noticeable deviation from $\beta_{\mathrm{EM-HM}}$ emerges in the nonlinear regime. In contrast, $\beta_{vv-vw+bb+jb}$, which includes contributions from turbulent magnetic energy \verb|spec_mag(k, t)| and current helicity \verb|k*k*spechel_mag(k, t)|, coincides with $\beta_{\mathrm{EM-HM}}$ in the kinematic regime and, remarkably, continues to agree with $\beta_{\mathrm{EM-HM}}$ even in the nonlinear regime. If helicity is removed in the forcing energy, these three $\beta$ profiles coincide.\\

$\alpha_{\text{MFT}}$, $\beta_{\text{MFT}}$, $\beta_{vv-vw}$, and $\beta_{vv-vw+bb+jb}$ were obtained using the code below. In principle, $\alpha$ and $\beta$ should be calculated using data from the turbulent regime; however, the precise range has not been definitively established. To minimize errors, the entire domain was divided into three scales: large scale, forcing scale, and dissipation scale \cite{2012MNRAS.419..913P}. Data within the forcing scale regime ($k=2\sim3$ to $k=9$) were then utilized. Figures A2(a) and A2(b) in the Appendix compare the reconstructed $\overline{B}$ fields with the DNS result for $k=2$–$8$ and $k=2$–$200$, respectively. Errors begin to emerge when the range is extended to $k>9$. This suggests that the $\alpha$ and $\beta$ effects are determined primarily by the forcing scale rather than the dissipation scale. Additionally, the non-memory effect of turbulence was considered in the time integration. If all past event history is included, the cumulative error expands to the point where measurements become meaningless.
\begin{verbatim}
for j=0, i_last do begin

    for i=2, 9 do begin
      vw_2_kmax[j]=vw_2_kmax[j]+spechel_kin(i, j)               ; kinetic helicity
      jb_2_kmax[j]=jb_2_kmax[j]+(k[i])^2*spechel_mag(i, j)      ; current helicity
      vv_2_kmax[j]=vv_2_kmax[j]+2.0*spec_kin(i, j)              ; kinetic energy
      bb_2_kmax[j]=bb_2_kmax[j]+2.0*spec_mag(i, j)              ; magnetic energy 
    endfor

    alpha_MFT[j]=(jb_2_kmax[j]-vw_2_kmax[j])/3.0                ; mean field alpha
    beta_MFT[j]=(vv_2_kmax[j])/3.0                              ; mean field beta
  endfor
\end{verbatim}
Here, \verb|spechel_kin(i, j)| indicates $H_V(k, t)(=\langle {\bf v}\cdot {\bf \omega} \rangle)$, and \verb|spec_kin(i, j)| indicates $E_V(k, t)(=\langle u^2 \rangle/2)$.\\

Figure \ref{f3a} illustrates the curl of the electromotive force $\nabla \times \langle {\bf u}\times {\bf b}\rangle$. The red solid line represents the direct output from the DNS data, whereas the other lines correspond to the $\alpha$, $\beta$ theoretical models. The DNS EMF was evaluated from the time rate of change of the large-scale magnetic field: $\partial \overline{B}/\partial t - \eta \nabla^2\overline{B}$. However, the other models were calculated directly based on their respective models. The black dotted line represents the EMF constructed from $\alpha_{{EM-HM}}$ and $\beta_{{EM-HM}}$, showing excellent agreement with the DNS results (red solid line). In contrast, the model using $\beta_{vv-vw}$ (black dashed line) aligns well only in the kinematic regime and diverges as the magnetic field strength grows. Furthermore, $\beta_{vv-vw+bb+jb}$ (blue line) is suppressed by the current helicity, thereby approaching the DNS results in the whole range. Except for the DNS data and $\alpha_{\text{EM-HM}}$ and $\beta_{\text{EM-HM}}$, a smoothing filter was applied to the remaining models; its necessity is readily apparent from their noisy profiles. Obtaining reliable results from such noisy data requires an appropriate theoretical model with suitable filtering techniques. However, the large-scale magnetic field yields smooth, almost noise-free data, rendering filtering unnecessary. This procedure is clearly demonstrated in the IDL source code provided below.
\begin{verbatim}
for j=0L,  i_last-1 do begin
    dB_dt[j]= ( Sqrt(2.0*spec_mag(1, j+1)) - Sqrt(2.0*spec_mag(1, j)) ) / $
              (tt_power_mag_time[j+1] - tt_power_mag_time[j]).   ; dB / dt
endfor

for j=0L, i_last-1 do begin
    left[j]=dB_dt[j] + eta*Sqrt(2.0*spec_mag(1, j))                ; dB/dt + \eta k*k*B (k=1)
    right[j]=(alpha[j]*fhm[j] - beta[j]) * Sqrt(2.0*spec_mag(1, j))
; fhm[j]=spechel_mag(1, j)/(2*spe_mag(1, j))                        ; fhm = k<A.B>/<B.B>
endfor
\end{verbatim}

In Figure \ref{f4}, the large-scale magnetic field was reconstructed using the models of $\alpha$, $\beta$, and IDL code is as follows:
\begin{verbatim}
B_theory_MFT[0] = sqrt(2.0*spec_mag(1, 0))
for j=0L, i_last-1 do begin
    B_theory_MFT[j+1]=B_theory_MFT[j] + $
      (alpha_MFT[j]*fhm[j]-beta_MFT[j]-eta)*B_theory_MFT[j]*$
      (tt_power_mag_time[j+1]-tt_power_mag_time[j])       ; dt = t[j+1] - t[j]
endfor
\end{verbatim}
Note that the initial conditions for each equation \verb|B_theory[0] = sqrt(2.0*spec_mag(1, 0))| were adopted directly from the initial values of the code. This is an inevitable step in solving differential equations. From the next step, the system uses its own calculated values. Also it should be noted that when applying the $\alpha$-effect, only the helical component of the magnetic field must be applied $\nabla \times \overline{\bf B}=\lambda \overline{\bf B}$. However, because the magnetic helicity ratio varies over time, the magnetic field is corrected by multiplying it by $f_{hm}$, which is defined as $f_{hm}=k\langle {\bf A}\cdot {\bf B}\rangle /\langle B^2\rangle$

\begin{equation}
\overline{B}_{j+1} = \overline{B}_j + \big(sign*\alpha_j*f_{hm,\,j} - \beta_j - \eta\big)*\overline{B}_j*(t_{j+1} - t_j).
\label{B_field_IDL}
\end{equation} 
Sign$=-1$ for positive helical forcing (southern hemisphere), Sign$=+1$ for negative helical forcing (nothern hemisphere): $\nabla\times \overline{B}=+\overline{B}$ at $k=1$. Helicity ratio $f_{hm}$ is represented by \verb|k*spechel_mag(1, t)/2*spec_mag(1, t)|. The time interval is $\Delta t \sim 0.2$ and the correlation length $l = 2\pi / 3$ are used. The corresponding wavenumber is $k = 3$, which is the average wavenumber between the large scale ($k = 1$) and the energy injection scale ($k = 5$). This result is quite interesting and warrants further investigation. However, we will not discuss the details at this point. \footnote{Changing `$l$' introduces errors. The most accurate results are obtained for $l \approx 2$–$2.09$. This value was initially identified through trial and error, and was later confirmed to correspond to a characteristic wavenumber $k \sim 3$.}\\

After confirming that $\alpha_{EM-HM}$ and $\beta_{EM-HM}$ accurately reproduce the reliable results, we directly tested their impacts on magnetic field amplification. In Figure~\ref{f5}, the $\beta_{\mathrm{EM-HM}}$ effect is fully applied, while the $\alpha_{\mathrm{EM\text{-}HM}}$ effect is modified (100\%, 0\%, -100\%). This figure shows that the $\alpha$ effect does not play a main role during the early (kinematic) stage, but it becomes more significant as the magnetic field grows and reaches saturation. On the other hand, in Figure~\ref{f6}, the $\alpha_{\mathrm{EM\text{-}HM}}$ effect is fully applied, and the $\beta_{\mathrm{EM-HM}}$ effect is reduced. The dot-dot-dot-dashed line shows the result with 50\% of the $\beta$ effect, and the red dashed line shows the case with only 15\% of it. When the $\beta$ contribution drops below around 10\%–15\%, the magnetic field stops growing. 

\begin{verbatim}
for j=0L, i_last-1 do begin
    B_theory_alpha_0[j+1]=B_theory_alpha_0[j] + $
      (weight1*alpha[j]-weight2*beta[j]-eta)*B_theory[j]*$
      (tt_power_mag_time[j+1]-tt_power_mag_time[j])
endfor
\end{verbatim}

These facts challenge the conventional understanding of the $\alpha$ and $\beta$ effects in dynamo. While the $\alpha$ effect was traditionally considered the main driver in magnetic dynamos, diffusion actually exerts a larger influence. However, since helicity characteristics are simultaneously transported via turbulent diffusion, disentangling electromagnetic amplification from diffusion is non-trivial. Refer to the section [Schematic Model of (Non-)Helical Field Structures] in Appendix.\\

\subsection{Application: From Turbulent Dynamo Theory to Particle Distributions}
Having established that the turbulent dynamo coefficients \(\alpha\) and \(\beta\) can reproduce the large-scale magnetic field, we next consider how this dynamically generated field enters the kinetic description of the plasma particles. Beyond macroscopic impacts, such as explaining magnetic field cycles in stars or planets, the $\alpha$ and $\beta$ effects can influence the distribution of plasma particles $f$, too. $f$ is essentially described by the kinetic Boltzmann model. The particle distribution is significantly affected by magnetic fields, in addition to collision effects $f_c$. In particular, the large-scale magnetic field determines the distribution characteristics of the system. A brief expression is as follows:
\begin{eqnarray}
&&\frac{\partial f}{\partial t}+{\bf v}\cdot \nabla f+\frac{q_a}{m_a}\big({\bf E}+{\bf v}\times {\bf B} \big)=\frac{\delta f_c}{\delta t}\nonumber\\
\Rightarrow &&\frac{\partial f}{\partial t}+{\bf v}\cdot \nabla f+\frac{q_a}{m_a}\bigg({\bf E}+{\bf v}\times \int^t [\alpha(t) \nabla\times {\bf \overline{B}_0}+(\beta(t)+\eta)\nabla^2{\bf \overline{B}_0}] dt \bigg)=\frac{\delta f_c}{\delta t}.
\label{kinetic_Boltzmann}
\end{eqnarray} 
While several applications can be considered—such as Debye shielding and astrophysical nucleosynthesis—these will be addressed in a future study (refer to \cite{2024PhRvD.109j3002P}).\\



\section{Summary and Discussion}
In a rotating plasma system, buoyancy and Coriolis forces generate kinetic helicity, which in turn leads to the production of magnetic helicity or current helicity.\\

To quantitatively assess the influence of helicities on magnetic diffusivity, we derived theoretical models for $\alpha$ and $\beta$ and applied DNS raw data to obtain time-evolving profiles of $\alpha$ and $\beta$. Using these data, we reconstructed the large scale magnetic field and compared it with the DNS result to validate $\alpha$ and $\beta$. We found that kinetic helicity reduces $\beta$, thereby amplifying the magnetic field. In contrast, current helicity and magnetic energy as well as kinetic energy enhance $\beta$, suppressing the magnetic field. Kinetic helicity plays a central role in amplifying the magnetic field in the whole range, while current helicity becomes dominant in the nonlinear regime, preventing uncontrolled growth and driving the system toward saturation. Even in regions where the molecular diffusivity $\eta$ is very small, current helicity, magnetic energy, and kinetic energy in the turbulent regime still constrain magnetic field growth.\\

Furthermore, by applying weighting factors to $\alpha$ and $\beta$, we directly verified their respective contributions to magnetic field amplification. Contrary to the predictions of classical models, the $\alpha$-effect primarily sustains magnetic saturation in the nonlinear regime, while the $\beta$-effect plays a decisive role in magnetic field amplification and maintenance across all regimes. This indicates that, in plasma systems composed of a large number of particles exhibiting fluid-like behavior and electric neutrality, the transport of kinetic helicity via turbulent diffusion is the main driver of magnetic field amplification, regulated by energy and current helicity. An investigation into the very early dynamics of $\alpha$ and $\beta$ effects leaves it as yet unclear which factor determines the sign (or handedness) of the large-scale magnetic field. This ambiguity arises because the $\alpha$ and $\beta$ formulations in this study fundamentally assume a homogeneous and isotropic system.\\

In this manuscript, we introduced two complementary approaches: one based on differential equations using large-scale magnetic field data, and the other based on integral equations using kinetic and magnetic data from the turbulent regime. Despite the differences in methodology and required input data, both approaches produced very similar results and successfully reproduced the DNS magnetic field. While our present study uses DNS data to investigate $\alpha$ and $\beta$, these models are, in principle, applicable to observational data as well. Finally, to demonstrate the implementation of the theoretical model, the IDL analysis code is included across the main body and Appendix.\\


\section*{Data Availability Statement}
The IDL source codes and DNS datasets analyzed during the current study are available in the GitHub repository, \url{https://github.com/pkiwan93/dynamo-dns-analysis}.

\clearpage

\appendix

\section{Supplementary Figures} \label{app:code}

\setcounter{figure}{0}
\renewcommand{\thefigure}{A\arabic{figure}}

\subsection{Schematic Model of (Non-)Helical Field Structures}

\begin{figure*}[htbp]
\centering
    \includegraphics[width=0.8\textwidth]{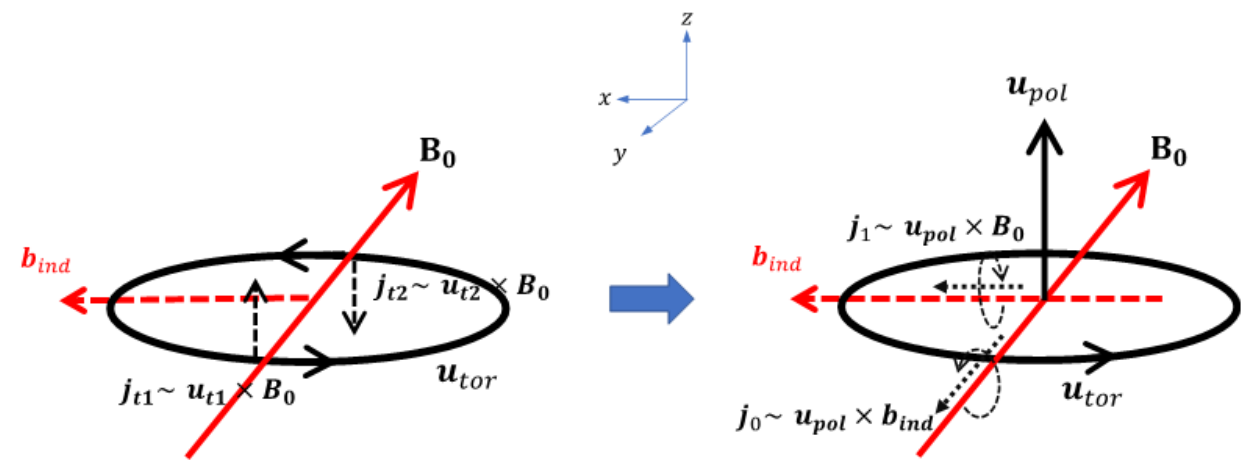}
\caption{Left: nonhelical field; right: helical field. \cite{particles8040098}}
\end{figure*}


Figure~A1 (left) demonstrates a circular plasma turbulent eddy embedded in a background magnetic field $\mathbf{B}_0$, which governs the eddy's geometry and energy dynamics. The plasma motions $\mathbf{u}_{t1}$ and $\mathbf{u}_{t2}$ react with $\mathbf{B}_0$ to generate anti-parallel current densities $\mathbf{j}_{t1} (\hat{z})$ and $\mathbf{j}_{t2} (-\hat{z})$. By Ampère's law, these current profiles give rise to an induced magnetic field $\mathbf{b}_{\text{ind}} (\hat{x})$, leading to magnetic diffusion through $\nabla \times \mathbf{j}_{t} \sim \nabla \times (\nabla \times \mathbf{b}) \sim -\nabla^2 \mathbf{b}$. Because this magnetic induction operates continuously, it eventually weakens the primary magnetic field. This series of events describes the turbulent magnetic diffusion $\beta$ effect stemming from velocity fluctuations, expressed as $\beta \sim \int \langle u^2 \rangle d\tau$, which corresponds to small-scale dynamo diffusion. Conversely, when buoyancy is introduced, a poloidal velocity field $\mathbf{u}_{\text{pol}}$ arises. This motion couples with $\mathbf{b}_{\text{ind}}$ to generate a current density $\mathbf{j}_{0} (\hat{y}) \sim \mathbf{u}_{\text{pol}} \times \mathbf{b}_{\text{ind}}$ aligned with $\mathbf{B}_0$, thereby exciting a toroidal magnetic field $\mathbf{b}_{\text{tor}}$ (dotted circle). The configuration of $\mathbf{B}_0$ and $\mathbf{b}_{\text{tor}}$ produces left-handed magnetic helicity $H_{M1}$ via the $\alpha$ effect. Simultaneously, the cross product of $\mathbf{u}_{\text{pol}}$ and $\mathbf{B}_0$ forms a current density $\mathbf{j}_{1}$, which pairs with $\mathbf{b}_{\text{ind}}$ to generate right-handed magnetic helicity $H_{M2}$. This sequential mechanism reveals that the activation of the $\alpha$ effect inherently requires prior magnetic diffusion $\beta$.\\

This discussion is originally from a brief concept in Krause's textbook\cite{1980opp..bookR....K}, which we introduced in \cite{particles8040098}. We explain the structure again due to its direct relevance. The key point is that the $\alpha$ effect is preceded by a diffusion process; that is, the transport of the medium or magnetic fields inherently carries the properties of (magnetic) helicity along with it.

\subsection{Supplementary Analysis of Dynamo Coefficients and Helicity Dynamics}
\begin{figure*}[htbp]
\centering
\subfigure[$\alpha$ in south]{
    \label{fig:app_a}
    \includegraphics[width=0.45\textwidth]{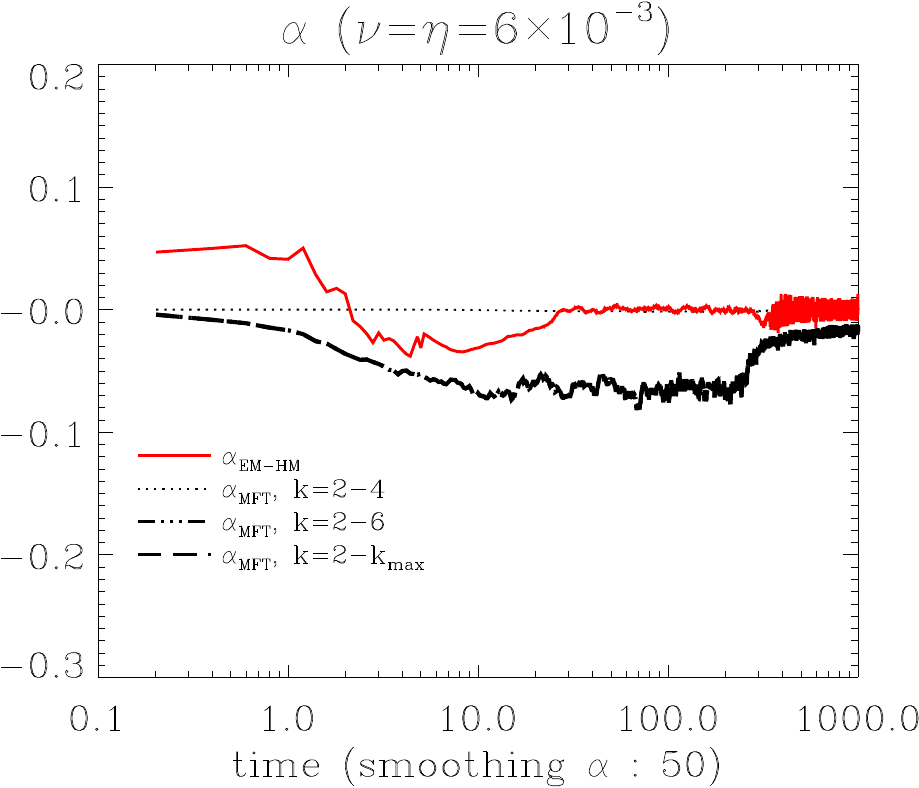}
}\hspace{-3mm}
\subfigure[$\beta$ in south]{
    \label{fig:app_b}
    \includegraphics[width=0.45\textwidth]{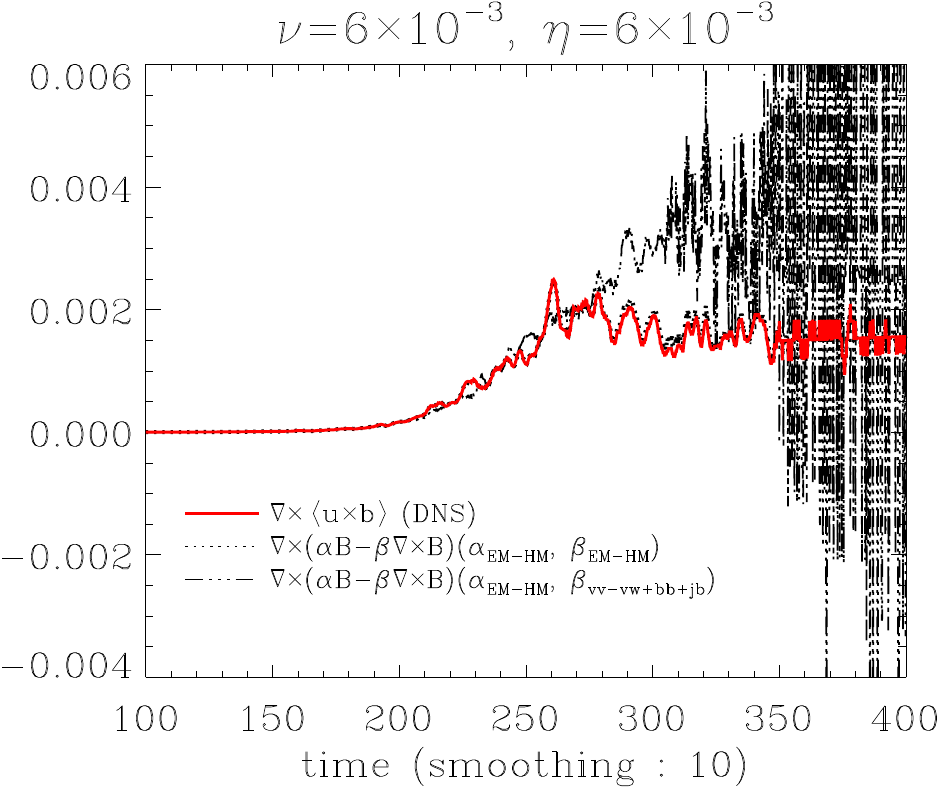}
}\\ \vspace{-3mm}
\subfigure[$f_h$ in south]{
    \label{fig:app_c}
    \includegraphics[width=0.45\textwidth]{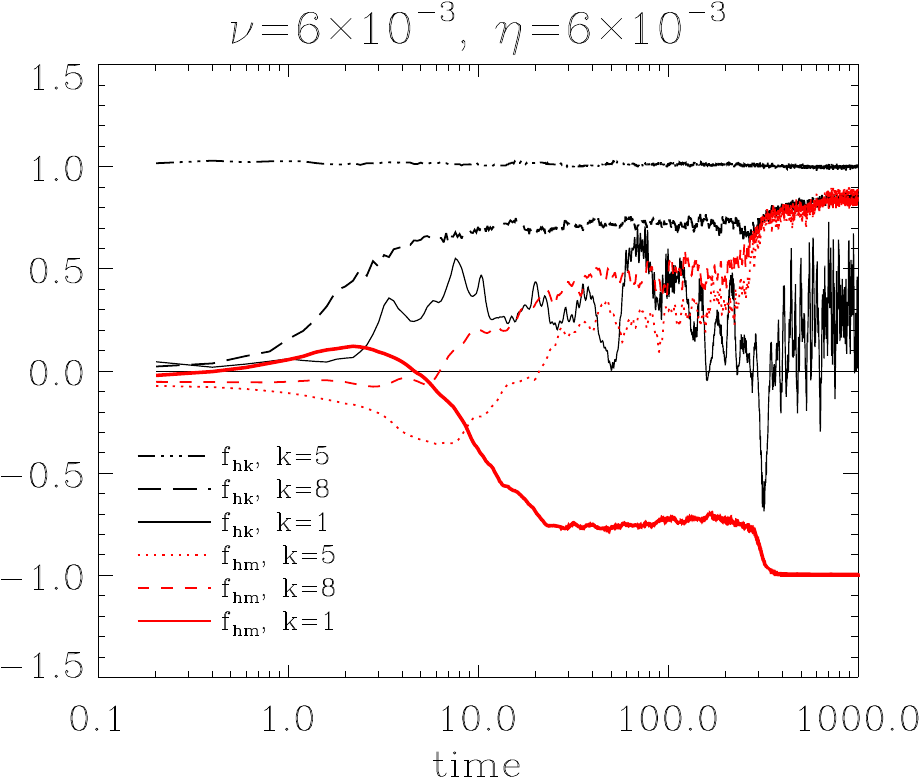}
}\hspace{-3mm}
\subfigure[$f_h$ in north]{
    \label{fig:app_d}
    \includegraphics[width=0.45\textwidth]{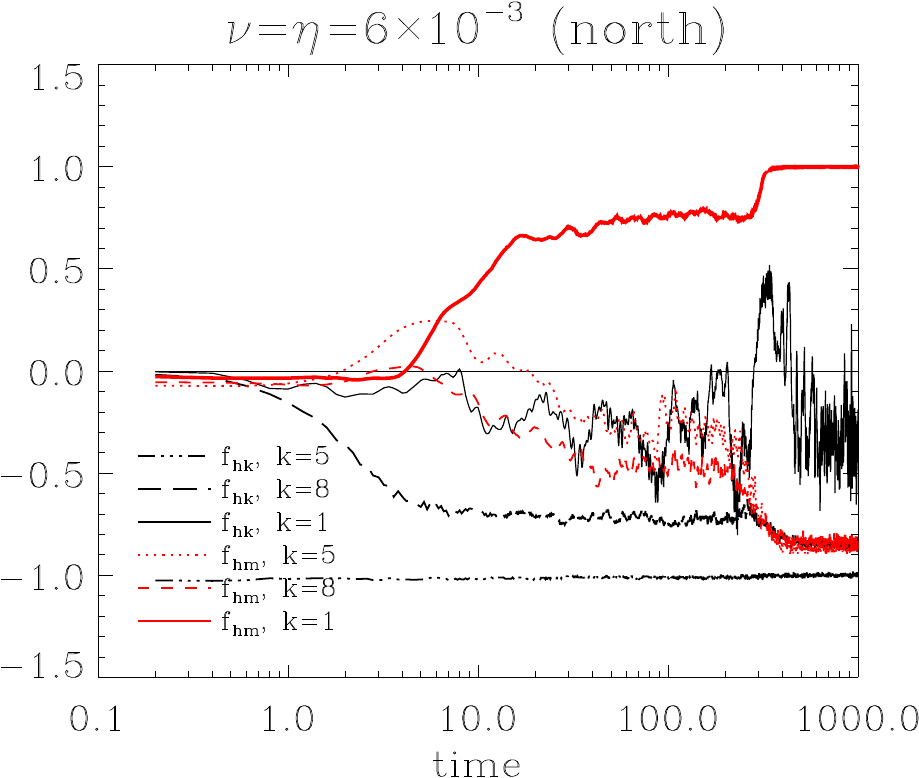}
}\\ \vspace{-3mm}
\subfigure[]{
    \label{fig:app_e}
    \includegraphics[width=0.45\textwidth]{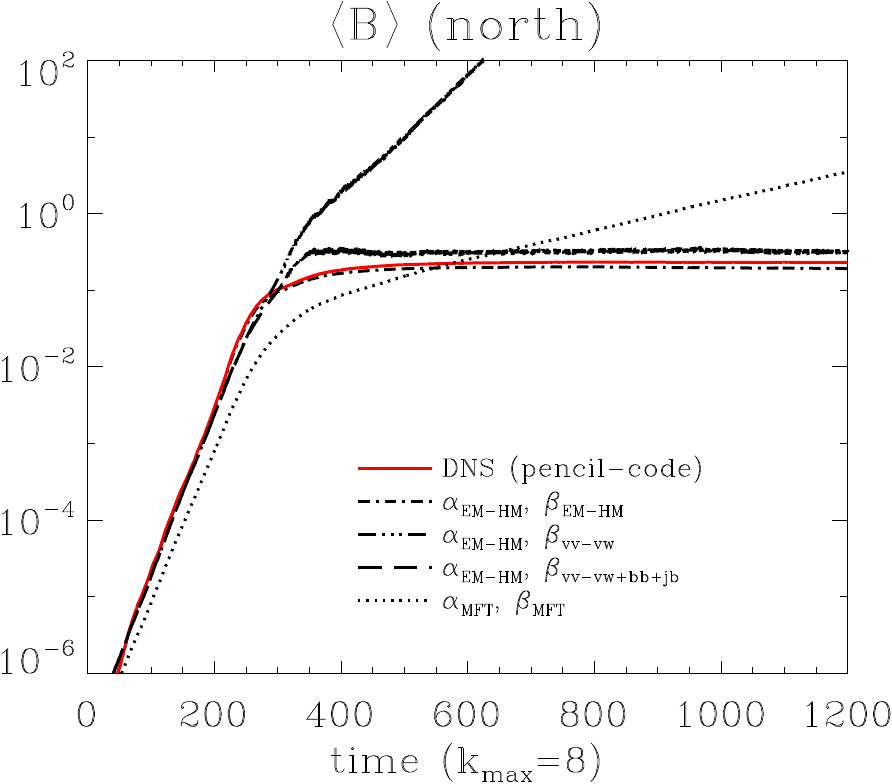}
}\hspace{-3mm}
\subfigure[]{
    \label{fig:app_f}
    \includegraphics[width=0.45\textwidth]{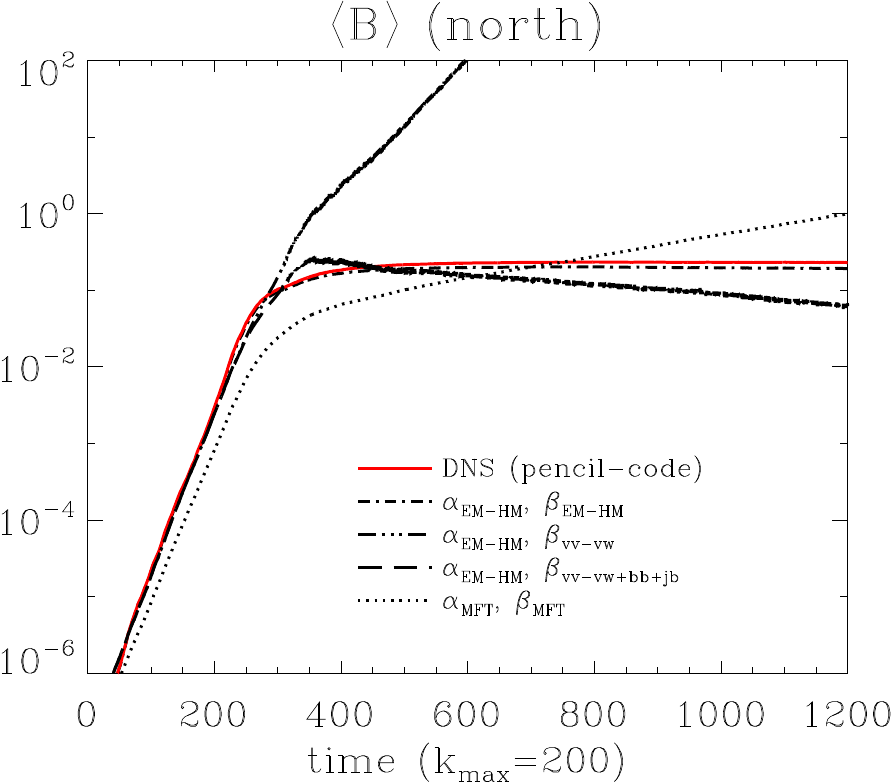}
}
\caption{Helicity ratio: $f_{hk} = H_V/2kE_V$, $f_{hm} = kH_M/E_M$. \textit{Reassembled and adapted from Ref.~\cite{particles8040098}.}}
\label{fig:app_spectrum}
\end{figure*}

Figure A1a shows the profile of $\alpha$ driven by positive kinetic helicity in the Southern Hemisphere. This corresponds to Figure \ref{f2} for the Northern Hemisphere, showing an opposite sign. The Southern Hemisphere $\beta$ is omitted because it is identical to that of the Northern Hemisphere.Figure  A1b corresponds to Figure \ref{f2}. Although the sign of $\alpha$ differs between hemispheres, taking the curl of the large-scale magnetic field also reverses the sign. Consequently, it generates a magnetic field of the same magnitude.Figures  A1c and  A1d show the kinetic helicity ratio $f_{hk}$ and the magnetic helicity ratio $f_{hm}$. Note that the helicity signs at the forcing scale ($k=5$) and the large scale are opposite.When analyzing data, we used the IDL smoothing function to average neighboring values. This accommodates the random input selection inherent to the Pencil Code (Figure A1b). In this paper, smoothing widths between 10 and 50 were selected. Other codes may use smaller or larger values, and FFT methods are also applicable.\\

Figures \ref{f7} and \ref{f8} correspond to Figures \ref{f5} and \ref{f6}, respectively, demonstrating the roles of the $\alpha$- and $\beta$-effects during the early stage of the system. They show that the $\beta$ effect remains dominant over the $\alpha$ effect. This dominance occurs not only because the magnitude of the $\alpha$ itself is smaller than $\beta$, but also because the $\alpha$ effect acts exclusively on the helical component of the large-scale magnetic field. As shown in Figures A1(c) and A1(d) in the Appendix, the magnetic helicity ratio $f_{hm}$ of the large-scale magnetic field at $k=1$ approaches $\pm 1$ over time. This effect is included in the code; omitting this correction leads to a slight increase in error within the kinematic regime. Meanwhile, the qualitative explanation of the dominance of the $\beta$ effect over the $\alpha$ effect was made in \cite{particles8040098} using a field structure model. This structure intuitively illustrates that before the $\alpha$ effect emerges, the magnetic field first undergoes a diffusion process, after which the helical component induces the electromagnetic $\alpha$ effect. In the absence of a helical component, a nonhelical system, the diffused magnetic field can no longer interact and eventually decays after fluctuating due to Ohmic dissipation.\\

Figure A1(e) includes the turbulent scales from $k=2$ to $k=8$, whereas Figure A1(f) covers $k=2$ to $k_{\text{last}}=200$. With the consideration of Figure \ref{f4}, $\alpha$ and $\beta$ coefficients are determined within the forcing scale regime ($k=2$–$8\sim9$), rather than over the entire turbulent scale.\\

\clearpage

\section{IDL code}
\begin{verbatim}[fontsize=\footnotesize]
pro spectrum_alpha_beta_north_PRD

; =====================================================================
; Code Title: IDL Routine for Alpha/Beta Analysis
; Author: Kiwan Park (Institute of Plasma Turbulence and Magnetic Fields)
; Description: Calculation of alpha and beta in the forcing scale regime.
; =====================================================================

  power_mag='power_mag_nu006_eta006_fh_1_hkfd.dat'                    ; reading Pencil-code data
  powerhel_mag='powerhel_mag_nu006_eta006_fh_1_hkfd.dat'
  power_kin='power_kin_nu006_eta006_fh_1_hkfd.dat'
  powerhel_kin='powerhel_kin_nu006_eta006_fh_1_hkfd.dat'

  Number_of_k=200                         ; Wavenumber
  eta=0.006                                    ; Molecular diffusivity

  spectrum_mag=dblarr(Number_of_k)           ; EM(k, t) = <B*B>/2
  spectrumhel_mag=dblarr(Number_of_k)       ; HM(k, t) = <A*B>
  spectrum_kin=dblarr(Number_of_k)             ; EV(k, t) = <U*U>/2
  spectrumhel_kin=dblarr(Number_of_k)         ; HV(k, t) = <U*W>

  k=findgen(Number_of_k)         

  ;;;;;;;;;;;;;;;;;;;;;;;;;;;;;;;;;;;;;;;;;;;;;;;;;;;;;;;;;
  close, 1
  i=0L
  openr, 1, power_mag           ; reading EM
  while not eof(1) do begin
    readf, 1, power_mag_time
    readf, 1, spectrum_mag
    i=i+1L
  endwhile
  close, 1

  tt_power_mag_time=dblarr(i)  ; Array of kinetic time 
  tt_power_kin_time=dblarr(i)   ; Magnetic time, same as kinetic time.
  alpha=dblarr(i)                    ; Alpha
  beta=dblarr(i)                     ; Beta

  alpha_MFT=dblarr(i)            ; Alpha in MFT
  beta_MFT=dblarr(i)             ; Beta in MFT

  c=dblarr(i)                       ; B.B + A.B
  d=dblarr(i)                       ; B.B - A.B

  left=dblarr(i)                    ; dB/dt - eta*Laplacian B 
  right=dblarr(i)                  ; (alpha*fhm-beta)*B
  right2=dblarr(i)                ; (alpha*fhm-beta_{vv-vw+bb+jb})*B

  vw_2_4=dblarr(i)              ; turbulent kinetic helicity form k=2 to k=4
  jb_2_4=dblarr(i)               ; turbulent current helicity
  vv_2_4=dblarr(i)               ; turbulent kinetic energy
  bb_2_4=dblarr(i)              ; turbulent magnetic energy
  beta_vv_vw_2_4=dblarr(i)   ; turbulent beta from kinetic turbulence
  beta_bb_jb_2_4=dblarr(i)   ; turbulent beta from magnetic turbulence

  vw_2_6=dblarr(i)
  jb_2_6=dblarr(i)
  vv_2_6=dblarr(i)
  bb_2_6=dblarr(i)
  beta_vv_vw_2_6=dblarr(i)
  beta_bb_jb_2_6=dblarr(i)

  vw_2_10=dblarr(i)
  vv_2_10=dblarr(i)
  bb_2_10=dblarr(i)
  jb_2_10=dblarr(i)
  beta_vv_vw_2_10=dblarr(i)
  beta_bb_jb_2_10=dblarr(i)

  vw_2_12=dblarr(i)
  vv_2_12=dblarr(i)
  bb_2_12=dblarr(i)
  jb_2_12=dblarr(i)
  beta_vv_vw_2_12=dblarr(i)
  beta_bb_jb_2_12=dblarr(i)

  vw_2_kmax=dblarr(i)
  jb_2_kmax=dblarr(i)
  vv_2_kmax=dblarr(i)
  bb_2_kmax=dblarr(i)
  beta_vv_vw_2_kmax=dblarr(i)
  beta_bb_jb_2_kmax=dblarr(i)

  dB_dt=dblarr(i)
  B_theory=dblarr(i)  ; theoretical calculation of B field
  B_theory2=dblarr(i)
  B_theory3=dblarr(i)
  B_theory_MFT=dblarr(i)

  B_theory_alpha_1=dblarr(i)
  B_theory_alpha_0=dblarr(i)
  B_theory_beta_1=dblarr(i)
  B_theory_beta_0=dblarr(i)


  EV=dblarr(i)
  kEV=dblarr(i)
  EM=dblarr(i)
  kEM=dblarr(i)

  spec_kin=dblarr(Number_of_k, i)   ; array for E_V
  spechel_kin=dblarr(Number_of_k, i) ; array for H_V

  spec_mag=dblarr(Number_of_k, i)  ; array for E_M
  spechel_mag=dblarr(Number_of_k, i) ; array for H_M

  i_last=i-1

  i=0L

  close, 1
  openr, 1, power_kin    ; reading EV data
  while not eof(1) do begin
    readf, 1, power_kin_time
    readf, 1, spectrum_kin
    tt_power_kin_time(i)=power_kin_time
    spec_kin(*, i)=spectrum_kin
    i=i+1L
  endwhile
  close, 1

  i=0
  ;;;;;;;;;;;;;;;;;;;;;;;;;;;;;;;;;;;;;;;;;;;;;;;;;;;;;;;;;;;;;;;;
  close, 1
  openr, 1, powerhel_kin
  while not eof(1) do begin
    readf, 1, powerhel_kin_time
    readf, 1, spectrumhel_kin
    tt_power_kin_time(i)=powerhel_kin_time
    spechel_kin(*, i)=spectrumhel_kin
    i=i+1L
  endwhile
  close, 1

  i=0L

  close, 1
  openr, 1, power_mag
  while not eof(1) do begin
    readf, 1, power_mag_time
    readf, 1, spectrum_mag
    tt_power_mag_time(i)=power_mag_time
    spec_mag(*, i)=spectrum_mag
    i=i+1L
  endwhile
  close, 1

  i=0

  close, 1
  openr, 1, powerhel_mag
  while not eof(1) do begin
    readf, 1, powerhel_mag_time
    readf, 1, spectrumhel_mag
    tt_power_mag_time(i)=powerhel_mag_time
    spechel_mag(*, i)=spectrumhel_mag
    i=i+1L
  endwhile
  close, 1

  for j=0L,  (i_last-1) do begin
    c[j]=( (2.0*spec_mag(1, j) + 0.99*spechel_mag(1, j)) )
    d[j]=( (2.0*spec_mag(1, j) - 0.99*spechel_mag(1, j)) )
 
  endfor
 
  a=0L

  for j=0L,  i_last-1 do begin   ; Eqs. (7), (8)
    alpha[j] =0.25*( (ALOG(c[j+1])-ALOG(c[j]) )-(ALOG(d[j+1])-ALOG(d[j]))  )/$
                 (tt_power_mag_time[j+1]-tt_power_mag_time[j])
    beta[j] =-0.25*( (ALOG(c[j+1])-ALOG(c[j]))+(ALOG(d[j+1])-ALOG(d[j]))  )/$
                 (tt_power_mag_time[j+1]-tt_power_mag_time[j])-eta
  endfor
 
  B_theory[0] =  sqrt(2.0*spec_mag(1, 0))

  for j=0L, i_last-1 do begin   
    B_theory[j+1]=B_theory[j] + (alpha[j]*spechel_mag(1, j)/(2.0*spec_mag(1, j))-beta[j]-eta)$
                       * B_theory[j] * (tt_power_mag_time[j+1]-tt_power_mag_time[j])
    ;print, 'j, B_theory[j]', j, tt_power_mag_time[j], B_theory[j]
  endfor


  ;  for Eqs. (10), (14)
  for j=0L, i_last do begin
    for i=2L, 4 do begin
      vw_2_4[j]=vw_2_4[j]+spechel_kin(i, j)
      jb_2_4[j]=jb_2_4[j]+(k[i])^2*spechel_mag(i, j)
      vv_2_4[j]=vv_2_4[j]+2.0*spec_kin(i, j)
      bb_2_4[j]=bb_2_4[j]+2.0*spec_mag(i, j)
    endfor
  endfor

  for j=0L, i_last do begin
    for i=2L, 6 do begin
      vw_2_6[j]=vw_2_6[j]+spechel_kin(i, j)
      jb_2_6[j]=jb_2_6[j]+(k[i])^2*spechel_mag(i, j)
      vv_2_6[j]=vv_2_6[j]+2.0*spec_kin(i, j)
      bb_2_6[j]=bb_2_6[j]+2.0*spec_mag(i, j)
    endfor
  endfor

  for j=0L, i_last do begin
    for i=2L, 10 do begin
      vw_2_10[j]=vw_2_10[j]+spechel_kin(i, j)
      jb_2_10[j]=jb_2_10[j]+(k[i])^2*spechel_mag(i, j)
      vv_2_10[j]=vv_2_10[j]+2.0*spec_kin(i, j)
      bb_2_10[j]=bb_2_10[j]+2.0*spec_mag(i, j)
    endfor
  endfor

  for j=0L, i_last do begin
    for i=2L, 12 do begin
      vw_2_12[j]=vw_2_12[j]+spechel_kin(i, j)
      jb_2_12[j]=jb_2_12[j]+(k[i])^2*spechel_mag(i, j)
      vv_2_12[j]=vv_2_12[j]+2.0*spec_kin(i, j)
      bb_2_12[j]=bb_2_12[j]+2.0*spec_mag(i, j)
    endfor
  endfor

  for j=0L, i_last do begin
    for i=2L, 9+0*(Number_of_k-1) do begin
      vw_2_kmax[j]=vw_2_kmax[j]+spechel_kin(i, j)
      jb_2_kmax[j]=jb_2_kmax[j]+(k[i])^2*spechel_mag(i, j)
      vv_2_kmax[j]=vv_2_kmax[j]+2.0*spec_kin(i, j)
      bb_2_kmax[j]=bb_2_kmax[j]+2.0*spec_mag(i, j)
    endfor
    alpha_MFT[j]=(jb_2_kmax[j]-vw_2_kmax[j])/3.0
    beta_MFT[j]=(vv_2_kmax[j])/3.0
  endfor

  l=2.0*3.14159/3.0

  for j=0L, i_last do begin
    beta_vv_vw_2_6[j]=(vv_2_6[j])/3.0+l*(vw_2_6[j])/6.0
    beta_vv_vw_2_10[j]=(vv_2_10[j])/3.0+l*(vw_2_10[j])/6.0
    beta_vv_vw_2_12[j]=(vv_2_12[j])/3.0+l*(vw_2_12[j])/6.0
    beta_vv_vw_2_kmax[j]=(vv_2_kmax[j])/3.0+l*(vw_2_kmax[j])/6.0
  endfor

  for j=0L, i_last do begin
    beta_bb_jb_2_6[j]=(bb_2_6[j])/3.0-l*(jb_2_6[j])/6.0
    beta_bb_jb_2_10[j]=(bb_2_10[j])/3.0-l*(jb_2_10[j])/6.0
    beta_bb_jb_2_12[j]=(bb_2_12[j])/3.0-l*(jb_2_12[j])/6.0
    beta_bb_jb_2_kmax[j]=(bb_2_kmax[j])/3.0-l*(jb_2_kmax[j])/6.0
  endfor

  for j=0L,  i_last-1 do begin     ; for Figure 2(a)
    dB_dt[j]= ( Sqrt(2.0*spec_mag(1, j+1)) - Sqrt(2.0*spec_mag(1, j)) ) / $
   (tt_power_mag_time[j+1] - tt_power_mag_time[j])
  endfor

  for j=0L, i_last-1 do begin  ; left_ : dB/dt + \eta B = curl of EMF , right_ : alpha B - \beta B
    left[j]=dB_dt[j] + eta*Sqrt(2.0*spec_mag(1, j))
    right[j]=(alpha[j]*spechel_mag(1, j)/(2.0*spec_mag(1, j)) - beta[j]) * Sqrt(2.0*spec_mag(1, j))
    right2[j]=(alpha[j]*spechel_mag(1, j)/(2.0*spec_mag(1, j)) - beta_vv_vw_2_kmax[j]$
    -1*beta_bb_jb_2_kmax[j]) * Sqrt(2.0*spec_mag(1, j))
  endfor

  B_theory2[0] = sqrt(2.0*spec_mag(1, 0))    ; Initial condition from the code, for Figure 2(b)
  B_theory3[0] = sqrt(2.0*spec_mag(1, 0))
  B_theory_MFT[0] = sqrt(2.0*spec_mag(1, 0))

  for j=0L, i_last-1 do begin  ; j <= 1915   ; for Figure 2(b), Eq. (16)
    B_theory2[j+1]=B_theory2[j] + $
      (alpha[j]*spechel_mag(1, j)/(2.0*spec_mag(1, j))-beta_vv_vw_2_kmax[j]-eta)*B_theory2[j]*$
      (tt_power_mag_time[j+1]-tt_power_mag_time[j])
  endfor

  for j=0L, i_last-1 do begin
    B_theory3[j+1]=B_theory3[j] + $
      (alpha[j]*spechel_mag(1, j)/(2.0*spec_mag(1, j))-beta_vv_vw_2_kmax[j]$
      -beta_bb_jb_2_kmax[j]-eta)*B_theory3[j]*$
      (tt_power_mag_time[j+1]-tt_power_mag_time[j])
  endfor

  for j=0L, i_last-1 do begin
    B_theory_MFT[j+1]=B_theory_MFT[j] + $
      (alpha_MFT[j]*spechel_mag(1, j)/(2.0*spec_mag(1, j))$
      -beta_MFT[j]-eta)*B_theory_MFT[j]*$
      (tt_power_mag_time[j+1]-tt_power_mag_time[j])
  endfor

  v=0.6
  h=0.25
  v1=-0.05

  set_plot, 'ps'

  device, filename='B_DNS_vs_B_theory_alpha_beta_comparison.ps'
  device, xsize=7, ysize=5.5, /inches
  device, decomposed=1, /color

  plot, tt_power_mag_time, B_theory,$
    title='!12<!6B!12>!6 (north)',  xtitle='!6time (!7D!6t=0.02, !8l!6=2!4p!6/3)',$ 
    xrange=[0.1, 1200], linestyle=3, thick=5, charsize=1.7, yrange=[1e-6, 1.2], /ylog
  oplot, tt_power_mag_time, sqrt(2.0*spec_mag(1, *)), linestyle=0, thick=5, color='ff'X
  oplot, tt_power_mag_time, B_theory2, linestyle=4, thick=5
  oplot, tt_power_mag_time, B_theory3, linestyle=5, thick=5
  oplot, tt_power_mag_time, B_theory_MFT, linestyle=1, thick=5

  h2=-0.05
  v2=0.7

  plots, [0.42-h2, 0.49-h2],  [0.32-v+v2, 0.32-v+v2], linestyle=0, thick=5, color='ff'X, /normal
  xyouts, 0.5-h2,  0.31-v+v2, 'DNS (pencil-code)', charsize=1.2, /normal  ;(!4V!650  X 50)

  plots, [0.42-h2, 0.49-h2],  [0.28-v+v2, 0.28-v+v2], linestyle=3, thick=5, /normal
  xyouts, 0.5-h2,  0.27-v+v2, '!7a!6!dEM-HM!n, !7b!6!dEM-HM!n', charsize=1.2, /normal 

  plots, [0.42-h2, 0.49-h2],  [0.24-v+v2, 0.24-v+v2], linestyle=4, thick=5, /normal
  xyouts, 0.5-h2,  0.23-v+v2, '!7a!6!dEM-HM!n, !7b!6!dvv-vw!n', charsize=1.2, /normal

  plots, [0.42-h2, 0.49-h2],  [0.20-v+v2, 0.20-v+v2], linestyle=5, thick=5, /normal
  xyouts, 0.5-h2,  0.19-v+v2, '!7a!6!dEM-HM!n, !7b!6!dvv-vw+bb+jb!n', charsize=1.2, /normal

  plots, [0.42-h2, 0.49-h2],  [0.16-v+v2, 0.16-v+v2], linestyle=1, thick=5, /normal
  xyouts, 0.5-h2,  0.15-v+v2, '!7a!6!dMFT!n, !7b!6!dMFT!n', charsize=1.2, /normal

  device, /close

  for j=0L, i_last-1 do begin
    B_theory_alpha_0[j+1]=B_theory_alpha_0[j] + $
      (-0*alpha[j]-beta[j]-eta)*B_theory[j]*$
      (tt_power_mag_time[j+1]-tt_power_mag_time[j])
  endfor

  for j=0L, i_last-1 do begin
    B_theory_alpha_1[j+1]=B_theory_alpha_1[j] + $
      (-1*alpha[j]-beta[j]-eta)*B_theory[j]*$
      (tt_power_mag_time[j+1]-tt_power_mag_time[j])
  endfor

  v=0.6
  h=0.25
  v1=-0.05

  set_plot, 'ps'

  device, filename='B_DNS_vs_B_theory_alpha_change.ps'
  device, xsize=7, ysize=5.5, /inches
  device, decomposed=1, /color

  plot, tt_power_mag_time, B_theory,$
    title='!12<!6B!12>!6 (north)',  xtitle='!6time', xrange=[0.1, 700],$
    linestyle=3, thick=5, charsize=1.7, yrange=[1e-6, 0.6], /ylog
  oplot, tt_power_mag_time, sqrt(2.0*spec_mag(1, *)), linestyle=0, thick=4, color='ff'X
  oplot, tt_power_mag_time, B_theory_alpha_0, linestyle=4, thick=4
  oplot, tt_power_mag_time, smooth(abs(B_theory_alpha_1), 10), linestyle=5, thick=4

  h2=0.0
  v2=0.65

  plots, [0.42-h2, 0.49-h2],  [0.32-v+v2, 0.32-v+v2], linestyle=0, thick=4, color='ff'X, /normal
  xyouts, 0.5-h2,  0.31-v+v2, 'DNS (pencil-code)', charsize=1.2, /normal  ;(!4V!650  X 50)

  plots, [0.42-h2, 0.49-h2],  [0.28-v+v2, 0.28-v+v2], linestyle=3, thick=4, /normal
  xyouts, 0.5-h2,  0.27-v+v2, '!7a!6!dEM-HM!n(100!7%!6), !7b!6!dEM-HM!n(100!7%!6)', $
  charsize=1.2, /normal  ;(!4V!650  X 50)

  plots, [0.42-h2, 0.49-h2],  [0.24-v+v2, 0.24-v+v2], linestyle=4, thick=4, /normal
  xyouts, 0.5-h2,  0.23-v+v2, '!7a!6!dEM-HM!n(0!7%!6), !7b!6!dEM-HM!n(100!7%!6)', $
  charsize=1.2, /normal  ;(!4V!650  X 50)

  plots, [0.42-h2, 0.49-h2],  [0.2-v+v2, 0.2-v+v2], linestyle=5, thick=4, /normal
  xyouts, 0.5-h2,  0.19-v+v2, '!7a!6!dEM-HM!n(-100!7%!6), !7b!6!dEM-HM!n(100!7%!6)', $
  charsize=1.2, /normal

  device, /close

  for j=0L,  i_last-1 do begin
    B_theory_beta_0[j+1]=B_theory_beta_0[j] + $
      (alpha[j]*spechel_mag(1, j)/(2.0*spec_mag(1, j))-0.5*beta[j]-eta)*B_theory[j]*$
      (tt_power_mag_time[j+1]-tt_power_mag_time[j])
  endfor

  for j=0L,  i_last-1 do begin
    B_theory_beta_1[j+1]=B_theory_beta_1[j] + $
      (alpha[j]*spechel_mag(1, j)/(2.0*spec_mag(1, j))-0.15*beta[j]-eta)*B_theory[j]*$
      (tt_power_mag_time[j+1]-tt_power_mag_time[j])
  endfor

  v=0.6
  h=0.25
  v1=-0.05

  set_plot, 'ps'

  device, filename='B_DNS_vs_B_theory_beta_change.ps'
  device, xsize=7, ysize=5.5, /inches
  device, decomposed=1, /color

  plot, tt_power_mag_time, B_theory,$
   title='!12<!6B!12>!6 (north)',  xtitle='!6time (!7D!6t=0.02, !8l!6=2!4p!6/3)', $
   xrange=[0.1, 800], linestyle=3, thick=5, charsize=1.7, /ylog, yrange=[1e-6, 0.6]
  oplot, tt_power_mag_time, sqrt(2.0*spec_mag(1, *)), linestyle=0, thick=5, color='ff'X
  oplot, tt_power_mag_time, B_theory_beta_0, linestyle=4, thick=5
  oplot, tt_power_mag_time, B_theory_beta_1, linestyle=5, thick=5

  h2=0.043
  v2=0.65

  plots, [0.42-h2, 0.49-h2],  [0.32-v+v2, 0.32-v+v2], linestyle=0, thick=4, color='ff'X, /normal
  xyouts, 0.5-h2,  0.31-v+v2, 'DNS (pencil-code)', charsize=1.2, /normal  ;(!4V!650  X 50)

  plots, [0.42-h2, 0.49-h2],  [0.28-v+v2, 0.28-v+v2], linestyle=3, thick=4, /normal
  xyouts, 0.5-h2,  0.27-v+v2, '!7a!6!dEM-HM!n(100!7%!6), !7b!6!dEM-HM!n(100!7%!6)', $
  charsize=1.2, /normal  ;(!4V!650  X 50)

  plots, [0.42-h2, 0.49-h2],  [0.24-v+v2, 0.24-v+v2], linestyle=4, thick=4, /normal
  xyouts, 0.5-h2,  0.23-v+v2, '!7a!6!dEM-HM!n(100!7%!6), !7b!6!dEM-HM!n(50!7%!6)', $
  charsize=1.2, /normal  ;(!4V!650  X 50)

  plots, [0.42-h2, 0.49-h2],  [0.2-v+v2, 0.2-v+v2], linestyle=5, thick=4, /normal
  xyouts, 0.5-h2,  0.19-v+v2, '!7a!6!dEM-HM!n(100!7%!6), !7b!6!dEM-HM!n(15!7%!6)',$
 charsize=1.2, /normal

  device, /close

  v=0.5
  h=0.25
  v1=-0.05

  set_plot, 'ps'

  device, filename='alpha_vs_jb_vw.ps'
  device, xsize=7, ysize=5.5, /inches
  device, decomposed=1, /color

  plot,   tt_power_mag_time, (jb_2_4-vw_2_4)/3.0,$
    title='!4a!6 (!7m!6=!7g!6=6!9X!610!u-3!n, north)', $  ;, f!dhk!n=1.0
    xtitle='!6time (smoothing !4a!6 : 50)', xrange=[0.1, 1000], linestyle=1, thick=3,$ 
    charsize=1.7, yrange=[-0.3, 0.2], /xlog

  oplot,  tt_power_kin_time, (jb_2_6-vw_2_6)/3.0, linestyle=4, thick=7
  oplot,  tt_power_kin_time, (jb_2_kmax-vw_2_kmax)/3.0, linestyle=5, thick=7
  oplot,  tt_power_kin_time, Smooth(alpha, 50), linestyle=0, thick=5, color='ff'X
 
  h2=0.17
  v2=1.15-0.53

  plots, [0.42-h2, 0.49-h2],  [0.32-v+v2, 0.32-v+v2], linestyle=0, thick=5, $
  color='ff'X, /normal
  xyouts, 0.5-h2,  0.31-v+v2, '!4a!6!dEM-HM!n', charsize=1.2, /normal  

  plots, [0.42-h2, 0.49-h2],  [0.28-v+v2, 0.28-v+v2], linestyle=1, thick=3, /normal
  xyouts, 0.5-h2,  0.27-v+v2, '!4a!6!dMFT!n, !6k=2-4', charsize=1.2, /normal  

  plots, [0.42-h2, 0.49-h2],  [0.24-v+v2, 0.24-v+v2], linestyle=4, thick=5, /normal
  xyouts, 0.5-h2,  0.23-v+v2, '!4a!6!dMFT!n, !6k=2-6', charsize=1.2, /normal

  plots, [0.42-h2, 0.49-h2],  [0.2-v+v2, 0.2-v+v2], linestyle=5, thick=5, /normal
  xyouts, 0.5-h2,  0.19-v+v2, '!4a!6!dMFT!n, !6k=2-k!dmax!n', charsize=1.2, /normal

  device, /close

  v=0.5
  h=0.25
  v1=-0.05

  set_plot, 'ps'

  device, filename='beta_vs_vv_vw_bb_jb.ps'
  device, xsize=7, ysize=5.5, /inches
  device, decomposed=1, /color

  plot, tt_power_mag_time, Smooth(beta, 50),$
    title='!4b!6 (!7m!6=!7g!6=6!9X!610!u-3!n, north)', $ ;, f!dhk!n=1.0
    xtitle='!6time (smoothing !4b!6 : 50)', xrange=[0.1, 1200], linestyle=0, thick=4,$
             charsize=1.7, /xlog, yrange=[-0.2, 0.1], /normal
  oplot,  tt_power_kin_time, Smooth(beta, 50), linestyle=0, thick=4, color='ff'X
  oplot,  tt_power_kin_time, (beta_vv_vw_2_kmax), linestyle=3, thick=5
  oplot,  tt_power_kin_time, (beta_vv_vw_2_kmax+beta_bb_jb_2_kmax), linestyle=5, $
            thick=5
  oplot,  tt_power_kin_time, beta_MFT, linestyle=1, thick=5
  oplot,  tt_power_kin_time, 0*vv_2_10, linestyle=0, thick=1

  h2=0.
  v2=1.15-0.55

  plots, [0.42-h2, 0.49-h2],  [0.32-v+v2, 0.32-v+v2], linestyle=0, thick=5, color='ff'X,  $
  /normal
  xyouts, 0.5-h2,  0.31-v+v2, '!4b!6!dEM-HM!n', charsize=1.2, /normal

  plots, [0.42-h2, 0.49-h2],  [0.28-v+v2, 0.28-v+v2], linestyle=3, thick=5, /normal
  xyouts, 0.5-h2,  0.27-v+v2, '!4b!6!dvv-vw!n (!6k=2-kmax)', charsize=1.2, /normal

  plots, [0.42-h2, 0.49-h2],  [0.23-v+v2, 0.23-v+v2], linestyle=5, thick=5, /normal
  xyouts, 0.5-h2,  0.22-v+v2, '!4b!6!dvv-vw+bb+jb!n (!6k=2-kmax)', charsize=1.2, $
  /normal

  xyouts, 0.84,  0.64, '!4b!6!dvv-vw+bb+jb!n', /normal
  xyouts, 0.84,  0.6, '!4b!6!dvv-vw!n', /normal

  plots, [0.42-h2, 0.49-h2],  [0.18-v+v2, 0.18-v+v2], linestyle=1, thick=3, /normal
  xyouts, 0.5-h2,  0.17-v+v2, '!4b!6!dMFT!n (!6k=2-k!dmax!n)', charsize=1.2, /normal

  device, /close

  v=0.6
  h=0.25
  v1=-0.05

  set_plot, 'ps'

  device, filename='checking_alpha_beta_new.ps'
  device, xsize=7, ysize=5.5, /inches
  device, decomposed=1, /color

  plot, tt_power_mag_time, Smooth(right, 20),$
    title='!7m!6=!7g!6=6!9X!610!u-3!n (north)',  xtitle='!6time (smoothing : 20)',$
    xrange=[100, 360], linestyle=1, thick=7, charsize=1.7, yrange=[-0.003, 0.005]
  oplot, tt_power_mag_time, Smooth(right2, 20), linestyle=4, thick=3
  oplot, tt_power_mag_time, Smooth(left, 20), linestyle=0, thick=5, color='ff'X

  h2=0.17
  v2=0.65

  plots, [0.42-h2, 0.49-h2],  [0.32-v+v2, 0.32-v+v2], linestyle=0, thick=4, /normal, color='ff'X
  xyouts, 0.5-h2,  0.31-v+v2, '!9GX !12<!6u!9X!6b!12> !6(DNS)', charsize=1.2, /normal

  plots, [0.42-h2, 0.49-h2],  [0.28-v+v2, 0.28-v+v2], linestyle=1, thick=3, /normal
  xyouts, 0.5-h2,  0.27-v+v2, '!9GX!6(!7a!6B-!7b!9GX!6B!6)(!7a!6!dEM-HM!n, !7b!6!dEM-HM!n)', $
  charsize=1.2, /normal  ;(!4V!650  X 50)

  plots, [0.42-h2, 0.49-h2],  [0.24-v+v2, 0.24-v+v2], linestyle=4, thick=3, /normal
  xyouts, 0.5-h2,  0.23-v+v2, '!9GX!6(!7a!6B-!7b!9GX!6B!6)(!7a!6!dEM-HM!n, !7b!6!dvv-vw+bb+jb!n)', $
  charsize=1.2, /normal
  
  device, /close

  v=0.6
  h=0.25
  v1=-0.05

  set_plot, 'ps'

  device, filename='Helicity_ratio.ps'
  device, xsize=7, ysize=5.5, /inches
  device, decomposed=1, /color

  plot,   tt_power_mag_time, spechel_kin(5, *)/(5*2.0*spec_kin(5, *)),$
   title='!7m!6=!7g!6=6!9X!610!u-3!n (north)', $ 
   xtitle='!6time', xrange=[0.1, 500], linestyle=4, thick=3, charsize=1.7, yrange=[-1.2, 1.2], /xlog
  oplot,  tt_power_kin_time, spechel_kin(8, *)/(16.0*spec_kin(8, *)), linestyle=5, thick=3
  oplot,  tt_power_kin_time, Smooth(spechel_kin(1, *)/(2.0*spec_kin(1, *)), 1), linestyle=0, thick=2
  oplot,  tt_power_kin_time, 0*spechel_kin(1, *), linestyle=0, thick=1

  oplot,  tt_power_kin_time, 5.0*spechel_mag(5, *)/(2.0*spec_mag(5, *)), linestyle=1, thick=3,$ 
  color='ff'X
  oplot,  tt_power_kin_time, 8.0*spechel_mag(8, *)/(2.0*spec_mag(8, *)), linestyle=2, thick=3, $
  color='ff'X
  oplot,  tt_power_kin_time, spechel_mag(1, *)/(2.0*spec_mag(1, *)), linestyle=0, thick=6, $
  color='ff'X

  h2=0.17
  v2=1.15-0.43

  plots, [0.42-h2, 0.49-h2],  [0.32-v+v2, 0.32-v+v2], linestyle=4, thick=3, /normal
  xyouts, 0.5-h2,  0.31-v+v2, 'f!dhk!n, k=5', charsize=1.2, /normal  ;(!4V!650  X 50)

  plots, [0.42-h2, 0.49-h2],  [0.28-v+v2, 0.28-v+v2], linestyle=5, thick=3, /normal
  xyouts, 0.5-h2,  0.27-v+v2, 'f!dhk!n, k=8', charsize=1.2, /normal  ;(!4V!650  X 50)

  plots, [0.42-h2, 0.49-h2],  [0.24-v+v2, 0.24-v+v2], linestyle=0, thick=3, /normal
  xyouts, 0.5-h2,  0.23-v+v2, 'f!dhk!n, k=1', charsize=1.2, /normal

  plots, [0.42-h2, 0.49-h2],  [0.2-v+v2, 0.2-v+v2], linestyle=1, thick=3, color='ff'X, /normal
  xyouts, 0.5-h2,  0.19-v+v2, 'f!dhm!n, k=5', charsize=1.2, /normal

  plots, [0.42-h2, 0.49-h2],  [0.16-v+v2, 0.16-v+v2], linestyle=2, thick=3, color='ff'X, /normal
  xyouts, 0.5-h2,  0.15-v+v2, 'f!dhm!n, k=8', charsize=1.2, /normal

  plots, [0.42-h2, 0.49-h2],  [0.12-v+v2, 0.12-v+v2], linestyle=0, thick=3, color='ff'X, /normal
  xyouts, 0.5-h2,  0.11-v+v2, 'f!dhm!n, k=1', charsize=1.2, /normal

  device, /close

 v=0.6
  h=0.25
  v1=-0.05

  set_plot, 'ps'
  device, filename='EmEv_spectrum_with_inset.ps'
  device, xsize=7, ysize=5.5, /inches
  device, decomposed=1, /color

  plot, k, spec_kin(*, 0), $
    title='E!dV!n, E!dM!n (!7m!6=!7g!6=6!9X!610!u-3!n, north)', $
    xtitle='!6k', xrange=[1,120], $
    linestyle=2, thick=2, charsize=1.7, yrange=[1E-17,1], /xlog, /ylog

  oplot, k, spec_kin(*,500), linestyle=2, thick=3
  oplot, k, spec_kin(*,1000), linestyle=2, thick=4
  oplot, k, spec_kin(*,7200), linestyle=2, thick=6

  oplot, k, spec_mag(*,0), linestyle=0, color='ff'X, thick=2
  oplot, k, spec_mag(*,500), linestyle=0, color='ff'X, thick=3
  oplot, k, spec_mag(*,1000), linestyle=0, color='ff'X, thick=4
  oplot, k, spec_mag(*,7200), linestyle=0, color='ff'X, thick=6
  oplot, k, spec_mag(*,5000), linestyle=0, color='ff'X, thick=8

  h2=0.1
  v2=0.65
  plots, [0.42-h2, 0.49-h2],  [0.28-v+v2, 0.28-v+v2], linestyle=0, color='ff'X, thick=3, /normal
  xyouts, 0.5-h2,  0.27-v+v2, '!6<E!dM!n>', charsize=1.2, /normal  ;(!4V!650  X 50)

  plots, [0.42-h2, 0.49-h2],  [0.24-v+v2, 0.24-v+v2], linestyle=2, thick=3, /normal
  xyouts, 0.5-h2,  0.23-v+v2, '!6<E!dV!n>', charsize=1.2, /normal  ;(!4V!650  X 50)
  xyouts, 0.42-h2,  0.17-v+v2, 't=0.2, 100, 200, 1440', charsize=1.2, /normal

  ;========================================================
  ;  Inset plot 
  ;========================================================
  plot, k, spec_kin(*,7200), $
    xrange=[1,12], yrange=[1E-7,1], $
    linestyle=2, thick=3, charsize=1.0, /ylog, $
    position=[0.6,0.6,0.90,0.90], $
    /noerase   

  oplot, k, spec_mag(*,7200), linestyle=0, color='ff'X, thick=3

  ok=where(k ge 8 and k le 9)
  oplot, k(ok), 0.015*(k(ok))^(-5.0/3), linestyle=0, thick=5
  xyouts, 0.80, 0.80, 'k!u-5/3!n', charsize=1.5, /normal

  ;========================================================
  device, /close

  v=0.6
  h=0.25
  v1=-0.05

  set_plot, 'ps'

  device, filename='Em_Hm_spectrum.ps'
  device, xsize=7, ysize=5.5, /inches
  device, decomposed=1, /color

  plot,   k, 2.0*spec_mag(*, 0), title='!7m!6=!7g!6=6!9X!610!u-3!n (north)', $  ;, f!dhk!n=1.0
    xtitle='!6k', xrange=[1, 20], linestyle=2, thick=2, charsize=1.7, yrange=[1E-17, 1], /xlog,$
    /ylog
  oplot,  k, 2.0*spec_mag(*, 15), linestyle=2, thick=3
  oplot,  k, 2.0*spec_mag(*, 500), linestyle=2, thick=4
  oplot,  k, 2.0*spec_mag(*, 7200), linestyle=2, thick=6

  oplot,  k, abs(spechel_mag(*, 0)), linestyle=0, color='ff'X, thick=2
  oplot,  k, abs(spechel_mag(*, 15)), linestyle=0, color='ff'X, thick=3
  oplot,  k, abs(spechel_mag(*, 500)), linestyle=0, color='ff'X, thick=4
  oplot,  k, abs(spechel_mag(*, 7200)), linestyle=0, color='ff'X, thick=6

  h2=-0.0
  v2=0.65
  plots, [0.42-h2, 0.49-h2],  [0.28-v+v2, 0.28-v+v2], linestyle=0, color='ff'X, thick=3, /normal
  xyouts, 0.5-h2,  0.27-v+v2, 'ABS(!12<!6A!9.!6B!12>!6)', charsize=1.2, /normal  ;(!4V!650  X 50)

  plots, [0.42-h2, 0.49-h2],  [0.24-v+v2, 0.24-v+v2], linestyle=2, thick=3, /normal
  xyouts, 0.5-h2,  0.23-v+v2, '!12<!6B!9.!6B!12>!6', charsize=1.2, /normal  ;(!4V!650  X 50)
  xyouts, 0.42-h2,  0.17-v+v2, 't=0.2, 3.2, 100, 1440', charsize=1.2, /normal

  device, /close

  v=0.6
  h=0.25
  v1=-0.05

  set_plot, 'ps'

  device, filename='Hv_Hm_spectrum.ps'
  device, xsize=7, ysize=5.5, /inches
  device, decomposed=1, /color

  plot,   k, spechel_kin(*, 0), title='H!dV!n, H!dM!n(!7m!6=!7g!6=6!9X!610!u-3!n)', $
    xtitle='!6k', xrange=[1, 20], linestyle=5, thick=0, charsize=1.7, /xlog, yrange=[-0.05, 0.05]
  oplot,  k, spechel_kin(*, 300), linestyle=5, thick=1
  oplot,  k, spechel_kin(*, 1500), linestyle=5, thick=2
  oplot,  k, spechel_kin(*, 1700), linestyle=5, thick=3
  oplot,  k, spechel_kin(*, 2500), linestyle=5, thick=4
  oplot,  k, spechel_kin(*, 7200), linestyle=5, thick=6

  oplot,  k, spechel_mag(*, 0), linestyle=0, color='ff'X, thick=2
  oplot,  k, spechel_mag(*, 300), linestyle=0, color='ff'X, thick=3
  oplot,  k, spechel_mag(*, 1500), linestyle=0, color='ff'X, thick=4
  oplot,  k, spechel_mag(*, 1700), linestyle=0, color='ff'X, thick=6
  oplot,  k, spechel_mag(*, 2500), linestyle=0, color='ff'X, thick=4
  oplot,  k, spechel_mag(*, 7200), linestyle=0, color='ff'X, thick=6

  h2=-0.2
  v2=0.65

  plots, [0.42-h2, 0.49-h2],  [0.28-v+v2, 0.28-v+v2], linestyle=0, color='ff'X, thick=3, /normal
  xyouts, 0.5-h2,  0.27-v+v2, 'H!dM!n', charsize=1.2, /normal  ;(!4V!650  X 50)
  plots, [0.42-h2, 0.49-h2],  [0.24-v+v2, 0.24-v+v2], linestyle=5, thick=3, /normal
  xyouts, 0.5-h2,  0.23-v+v2, 'H!dV!n', charsize=1.2, /normal  ;(!4V!650  X 50)
  xyouts, 0.42-h2,  0.18-v+v2, 't=0.2-1440', charsize=1.2, /normal

  device, /close

  v=0.6
  h=0.25
  v1=-0.05

  set_plot, 'ps'

  device, filename='Hv_Hc_spectrum.ps'
  device, xsize=7, ysize=5.5, /inches
  device, decomposed=1, /color

  plot,   k, k^2*spechel_mag(*, 0), title='!7m!6=!7g!6=6!9X!610!u-3!n (north)', $
    xtitle='!6k', xrange=[1, 20], linestyle=0, thick=0, charsize=1.7, /xlog, yrange=[-0.05, 0.05]

  oplot,  k, k^2*spechel_mag(*, 0), linestyle=0, thick=1, color='ff'X
  oplot,  k, k^2*spechel_mag(*, 500), linestyle=0, thick=1, color='ff'X
  oplot,  k, k^2*spechel_mag(*, 1500), linestyle=0, thick=2, color='ff'X
  oplot,  k, k^2*spechel_mag(*, 1700), linestyle=0, thick=3, color='ff'X
  oplot,  k, k^2*spechel_mag(*, 2500), linestyle=0, thick=4, color='ff'X
  oplot,  k, k^2*spechel_mag(*, 7200), linestyle=0, thick=6, color='ff'X

  oplot,  k, spechel_kin(*, 0), linestyle=5, thick=2
  oplot,  k, spechel_kin(*, 500), linestyle=5, thick=3
  oplot,  k, spechel_kin(*, 1500), linestyle=5, thick=4
  oplot,  k, spechel_kin(*, 1700), linestyle=5, thick=6
  oplot,  k, spechel_kin(*, 2500), linestyle=5, thick=4
  oplot,  k, spechel_kin(*, 7200), linestyle=5, thick=6

  h2=-0.2
  v2=0.65

  plots, [0.42-h2, 0.49-h2],  [0.28-v+v2, 0.28-v+v2], linestyle=0, color='ff'X, thick=3, /normal
  xyouts, 0.5-h2,  0.27-v+v2, '<!6J!9.!6B!12>!6', charsize=1.2, /normal  ;(!4V!650  X 50)
  plots, [0.42-h2, 0.49-h2],  [0.24-v+v2, 0.24-v+v2], linestyle=5, thick=3, /normal
  xyouts, 0.5-h2,  0.23-v+v2, '!12<!6v!9.!7x!12>!6', charsize=1.2, /normal  ;(!4V!650  X 50)
  xyouts, 0.42-h2,  0.18-v+v2, 't=0.2-1440', charsize=1.2, /normal

  device, /close

end
\end{verbatim}

\clearpage

\section*{Acknowledgements}
The author acknowledges the support from IPTM.

\bibliography{bibfile_2025_1222}
\bibliographystyle{apsrev4-2}

\label{lastpage}
\end{document}